\pdfoutput=1

\documentclass{aastex63}

\shorttitle{Multivariate Gaussian process Models For Stellar Activity}
\shortauthors{Gilbertson et al.}
\usepackage{hyperref}
\usepackage{graphicx}
\usepackage{amsmath}
\usepackage{gensymb}
\usepackage[frozencache=true, cachedir=.]{minted} % use created minted files
\usepackage{xcolor}
\definecolor{cjg}{rgb}{0.0, 0.5, 0.5}
\definecolor{ebf}{rgb}{0.4, 0.0, 0.6}

\newcommand{\glom}{{\tt GLOM} }
\newcommand{\glomns}{{\tt GLOM}}
\newcommand{\glomjl}{{\tt GPLinearODEMaker.jl} }
\newcommand{\glomjlns}{{\tt GPLinearODEMaker.jl}}

\defcitealias{Jones2017}{J17}
\defcitealias{Gilbertson2020}{G20}

\begin{document}

\correspondingauthor{Christian Gilbertson}
\email{cjg66@psu.edu}

\title{Toward Extremely Precise Radial Velocities: II. A Tool For Using Multivariate Gaussian processes to Model Stellar Activity}
% Using multivariate Gaussian processes to discover planets in spotty spectra

\author[0000-0002-1743-3684]{Christian Gilbertson}
\affiliation{Department of Astronomy \& Astrophysics, The Pennsylvania State University, 525 Davey Laboratory, University Park, PA 16802, USA}
\affiliation{Center for Exoplanets \& Habitable Worlds, The Pennsylvania State University, 525 Davey Laboratory, University Park, PA 16802, USA}
\affiliation{Institute for Computational and Data Sciences, The Pennsylvania State University, 525 Davey Laboratory, University Park, PA 16802, USA}
\affiliation{Penn State Astrobiology Research Center, University Park, PA 16802, USA}

\author[0000-0001-6545-639X]{Eric B. Ford}
\affiliation{Department of Astronomy \& Astrophysics, The Pennsylvania State University, 525 Davey Laboratory, University Park, PA 16802, USA}
\affiliation{Center for Exoplanets \& Habitable Worlds, The Pennsylvania State University, 525 Davey Laboratory, University Park, PA 16802, USA}
\affiliation{Institute for Computational and Data Sciences, The Pennsylvania State University, 525 Davey Laboratory, University Park, PA 16802, USA}
\affiliation{Penn State Astrobiology Research Center, University Park, PA 16802, USA}
\affiliation{Institute for Advanced Study, 1 Einstein Drive, Princeton, NJ 08540, USA}

\author[0000-0002-8152-6818]{David E. Jones}
\affiliation{Department of Statistics, Texas A\&M University, 3143 Blocker Building, College Station, TX 77843, USA}

\author[0000-0002-9761-4353]{David C. Stenning}
\affiliation{Department of Statistics \& Actuarial Science, Simon Fraser University, Burnaby, BC V5A 1S6, Canada}

% \author{Add yourself here}
%\affiliation{}

\date{\today}

\begin{abstract}
The radial velocity method is one of the most successful techniques for the discovery and characterization of exoplanets. 
Modern spectrographs promise measurement precision of ~0.2-0.5 m/s for an ideal target star. 
However, the intrinsic variability of stellar spectra 
%(caused by starspots, pulsations, convective motions, granulation, etc.) 
can mimic and obscure true planet signals at these levels. 
%A data-driven approach for detecting planetary RV signals amidst stellar activity has recently been proposed by \citet{Rajpaul2015} and refined by \citet{Jones2017}. 
%This approach uses a physically motivated multivariate Gaussian process (GP) to jointly model the apparent RV and multiple indicators of stellar activity, allowing the planetary RV component to be separated from the total RV signal. 
\citet{Rajpaul2015} and \citet{Jones2017} proposed applying a physically motivated, multivariate Gaussian process (GP) to jointly model the apparent Doppler shift and multiple indicators of stellar activity as a function of time, so as to separate the planetary signal from various forms of stellar variability. 
These methods are promising, but performing the necessary calculations can be computationally intensive and algebraically tedious. 
In this work, we present a flexible and computationally efficient software package, \glomjlns, for modeling multivariate time series using a linear combination of univariate GPs and their derivatives.
The package allows users to easily and efficiently apply various multivariate GP models and different covariance kernel functions. 
%(like those described in \citet{Jones2017}).
We demonstrate \glomjl by applying the \citet{Jones2017} model to fit measurements of the apparent Doppler shift and activity indicators derived from simulated active solar spectra time series affected by many evolving starspots. % and find the kernels best suited toward modelling this type of activity.
We show how \glomjl makes it easy to explore the effect of different choices for the GP kernel.
% The activity-induced apparent RV perturbation is not a one-to-one function of current spectroscopic indicators. 
% While this limits other approaches that attempt to correct the RV for each spectrum individually, our statistical framework combines spectroscopic and temporal information to reconstruct the apparent RV perturbation and improve the sensitivity of low-mass planets. 
We find that local kernels could significantly increase the sensitivity and precision of Doppler planet searches relative to the widely used quasiperiodic kernel.
\end{abstract}

\keywords{Exoplanet detection methods (489), Stellar activity (1580), Astronomy software (1855), Gaussian processes regression (1930), Time series analysis (1916)}
% Keywords changed recently, now pick six from http://astrothesaurus.org/concept-select/

% \tableofcontents
% 2. math section
% - model desc
% - optimization
% 3. Package Implementation
% - Julia is cool
% - covariance construction
% - Symbolic computation of kernel derivatives

% 4. Example application to J17 model
% 4.1.
% - figure 1, including J17 model equations version we are using, , means, measurement uncertainties, number of observations,... fitting and keplerian mean func
% 4.2. Introduce idea of efficiency comparing kernels
% - different kernels
% 4.3. Results for differnt kernels
% - evidence ratios and model comparison

% 5. Summary

\section{Introduction} \label{sec:intro}

Recent and upcoming stabilized spectrographs are pushing the frontier of the precision and stability of Doppler spectroscopy with the goal of detecting and characterizing low-mass planets around main-sequence stars. 
The specifications for these extremely precise radial velocity (EPRV) instruments are so impressive that intrinsic stellar variability is expected to limit their Doppler precision for most target stars. 
There are many physical mechanisms that contribute to stellar variability, such as pulsations, granulation, starspots, faculae, and long-term magnetic cycles. 
Stellar variability due to starspots, faculae and other rotationally-linked phenomena are particularly concerning, as the stellar rotation period is often included in the range of potential planet orbital periods. 
In multiple cases, stellar variability has led to claims of exoplanet discoveries \cite[e.g.,][]{Vogt2010,Dumusque2012,AngladaEscude2014} 
that have subsequently been retracted or called into question \cite[e.g.,][]{FerozHobson2014,Robertson2014a,Robertson2014b,Robertson2015,Rajpaul2016} upon further analysis of the spectroscopic time series. 
In order to robustly detect and accurately characterize low-mass planets via Doppler planet surveys, the exoplanet community must develop statistical models capable of jointly modeling planetary perturbations and intrinsic stellar variability. 

In response to this challenge, we present the \glomjl \added{(\glomns)} software package that provides a %powerful and flexible model and 
set of computational tools to facilitate the analysis of spectroscopic time series. 
The package is built on a powerful and flexible model \deleted{(\glomns)} for jointly analyzing the apparent Doppler shift and one or more spectroscopic indicators that serve as diagnostics of the stellar variability. %accounting for a wide variety of forms of stellar variability and spectroscopic diagnostics. 
%
%In order to realize their full potential, astronomers must develop new strategies for distinguishing true Doppler shifts from intrinsic stellar variability. 
%
The statistical model described here closely follows models originally proposed by \citet{Rajpaul2015} and \citet{Jones2017}, hereafter \citetalias{Jones2017}, for application to EPRV exoplanet surveys. 
However, the statistical model and codes are quite general, and we envision that this model could be valuable for a wide variety of multivariate time series problems in astronomy (e.g., analysis of multiband photometric time series of variable stars, measuring time delay for gravitationally lensed quasar images; see \citet{HuTak2020}) and likely other disciplines (e.g., econometrics, fMRIs). 
Therefore, we present the statistical model (\S\ref{sec:statmodel}), implementation (\S\ref{sec:implentation}) and code (Appendix \ref{sec:samplecode}) in a general way, so as to encourage the application of the \glom model across multiple disciplines. 
In \S\ref{sec:ex_app}, we validate the model and codes for application to EPRV planet surveys by analyzing synthetic data sets generated to simulate spectroscopic variability due to sunspot regions on the solar spectra \added{and compare the performance of several GP kernel functions for this use case}.
Of course, researchers from other disciplines should consider the appropriateness of the \glom model for their scientific goals prior to application to problems other than EPRV planet surveys.  
Finally, we discuss potential future improvements to the \glomjl package in \S\ref{sec:discussion}.
The appendices contain a summary of priors used for kernel hyperparameters (\S\ref{sec:priors}) and planetary parameters (\S\ref{sec:keppriors})
\replaced{. Example code is available in appendix \ref{sec:samplecode} and }{and an illustrative code example (\S\ref{sec:samplecode}).} \deleted{the package repository at
\href{https://github.com/christiangil/GPLinearODEMaker.jl}{https://github.com/christiangil/GPLinearODEMaker.jl}.}

\section{Statistical Model} \label{sec:statmodel}

The \glom model provides a likelihood for modeling multivariate time series as linear combinations of univariate Gaussian processes (GPs) and their derivatives. 
In this section, we provide a quick introduction to Gaussian process regression (GPR; \S\ref{sec:gps}) and multivariate GPR (\S\ref{sec:mvgps}), the details of the general \glom model (\S\ref{sec:model}), and useful expressions for \deleted{efficiently} efficient inference with the \glom model (\S\ref{sec:optimize}). 
 
\subsection{Gaussian process Regression (GPR)} \label{sec:gps}

A Gaussian process is a continuous, stochastic process such that every finite linear combination of the random variables it describes is normally distributed. 
It is useful to recognize that any linear function of a GP results in another GP, as we will use this property to our advantage below. 
GPR, or kriging, is the process of using a GP to estimate the relationship between a set of dependent and independent variables.
For example, in \citetalias{Jones2017}, the independent variable is the observation time and the dependent variables are measurements derived from stellar spectra (i.e. apparent radial velocities and (RVs) stellar variability indicators).
GPR provides a ``non-parametric'' way to model stochastic functions of unknown functional form. 
This makes GPR particularly useful for modeling correlated noise when the underlying noise model is poorly understood or so complex that it is computationally impractical to model from first principles. 
For example, Doppler exoplanet surveys typically result in residuals greater than expected based purely on photon noise. In early surveys, observations were typically so far apart in time, that one could model the excess scatter (often referred to as ``jitter'') as uncorrelated noise \citep[e.g.,][]{Ford2006}. 
However, modern Doppler exoplanet surveys often include many closely spaced observations that can then include coherent signals from stellar activity. In these cases, modeling the unmodeled stellar variability as a GP (rather than uncorrelated jitter) significantly improves robustness for detecting and characterizing exoplanets \citep{Dumusque2017}. 
%
%, both astrophysical 
%in data meant to detect exoplanets, both from astrophysical and instrumental sources.
% In this case, we use them to model how the current activity state of stars (approximated by activity indicators) are affecting the radial velocities that we are measuring from the spectra.
% that enters observations

A GP is fully described by a mean function and a covariance matrix or kernel function.
While the mean function is often taken to be zero, it can be set to any function. 
For example, the mean function for the apparent RV of a star could be set to the Keplerian RV signal due to a putative planet (with orbital parameters that are known, need to be estimated from the data, or some combination). 
The kernel function describes the covariance between any pair of the points drawn from a given GP.
Most kernel functions assume that the covariance between points is a function of absolute separation in the independent variable, often time, i.e. $|t-t'|$, and a set of hyperparameters (e.g., the amplitude, and time scales of the correlations).
Potential choices for kernel functions are discussed in \S\ref{sec:kernel}.

\subsection{Multivariate GPR} \label{sec:mvgps}

Normally, Gaussian processes are used to model a single output, but if one expects the outputs to be correlated, one can build a multivariate GP informed by the relationship between the various outputs.
The simplest case would be to assume independent GPs for each output. 
However, this approach is likely to result in a model that is more flexible than appropriate if there are underlying physical correlations between the outputs (relative to their mean functions). 
An overly flexible model is likely to result in unnecessarily broad posterior distributions and reduced marginal likelihood. 
In the context of EPRV exoplanet surveys, this would correspond to reduced statistical power for detecting planets and reduced precision of planet masses and orbital parameters. 

One approach for defining GPs with multiple outputs is to represent each output as a linear function of a smaller number of latent univariate GPs. 
This approach is commonly used in geostatistics, where it is known as ``cokriging'' \citep{Rasmussen2006}. 
Unfortunately, a simple linear combination of GPs is unlikely to be well suited to modeling stellar variability due to starspots, faculae and other rotationally-linked phenomena. 
To understand this, it is useful to consider a simplistic model for a starspot that simply removes flux (equally across all wavelengths) coming from affected patch of the star. 
If the star's rotation axis is in the plane of the sky and starspot is located at the center of the disk, then it is expected to affect the depth and width of spectral lines, but, by symmetry, there can be no perturbation to the apparent radial velocity. In contrast, a spot close to the limb of the star (and near the equatorial plane) is expected to affect the shape of each spectral lines less (due to a smaller projected area), but its rotation velocity would cause an asymmetric change in the line shape that results in an apparent Doppler shift. Of course, active regions are more complex (e.g., dark spots are accompanied by brighter facular regions and also inhibit the convective blueshift effect). Nevertheless, this demonstrates why one expects that modeling the effects of stellar activity will require a more sophisticated covariance function than that resulting from a simple linear combinations of GPs.

Just as one can construct each output of a multivariate GP as a linear combination of latent univariate GPs, one can construct a GP using other linear operations on latent GPs such as differentiation and convolution.
% TODO? we could justify why we look at derivatives here
In particular, the \glom model considers each GP output to be proportional to a linear combination of a latent GP and its time derivatives. 
This allows observations of each GP output to be contribute to learning the behavior of one shared latent GP. 
Such an approach was proposed for jointly modeling stellar photometry and Doppler observations by \citet{Aigrain2012} and adapted for application to Doppler observations alone by \citet{Rajpaul2015}. 

The multivariate GP model was generalized by \citetalias{Jones2017} to allow for arbitrary linear combinations of the latent GP and its derivatives and provided a practical approach to choosing which coefficients should be allowed to deviate from zero in the context of Doppler planet surveys. 
The simplistic spot model from the paragraph above can also help motivate such covariance functions. 
A spot has the largest effect on the observed fluxes when it is nearest disk center, but it can not induce an apparent Doppler shift when it is along the rotation axis.
While a spot is on the side of the star rotating toward (away from) the observer, it suppresses light from the blue (red) side of each line, causing an apparent redshift (blueshift). 
As a spot starts to rotates into view (or is about to rotate out of view), its effect is limited by both geometric effects and limb darkening. 
Therefore, the apparent Doppler shift is expected to be strongly correlated with the time derivative of the total amount of stellar activity on the visible side of the star.
The expectation is supported by data-driven analyses of simulated solar observations \citepalias{Jones2017}.
Additionally, analysis of HARPS-N solar observations reveals a temporal lag ($\simeq~7$ days) between traditional activity indicators and perturbations to the apparent radial velocity due to stellar activity\citep{CollierCameron2019,Thompson2020}. 
Taking a weighted sum of a function and its derivatives can effectively approximate a time lagged version of the function (for lags small compared to the timescale on which the function is changing). 
%Therefore, the \glom family of models appears promising for analyzing stellar variability. 

While powerful, modeling multiple outputs as a linear combination of a univariate GP and its derivatives can be complicated and expensive to implement.
Since the derivatives of the latent GP are not independent of the latent GP, one must calculate cross terms between the latent GP and its derivatives for each time delay.
(These terms would cancel out if each output were a linear sum of independent GPs.) 
Additionally, using traditional algorithms, the computational complexity for GPR is $\mathcal{O}((m\times~n)^3)$ where $n$ is the number of observations and $m$ is the number of outputs.

To help address these issues, we have created an open-source software package, %\glomjlns\footnote{\href{https://github.com/christiangil/GPLinearODEMaker.jl}{https://github.com/christiangil/GPLinearODEMaker.jl}} 
\glomjlns,
that allows users to easily calculate the likelihood for 
\glom (i.e., \citetalias{Jones2017}-type) models.
While the \glom model does not reduce the order of complexity, \glomjl mitigates the cost by providing a computationally and memory efficient implementation of the necessary algorithms. 
\glomjl also provides analytical gradients and Hessians for for \glom likelihood using standard or custom-created kernels.
This enables the use of geometric optimization and sampling algorithms to converge orders of magnitude faster than algorithms using the likelihood alone.
Additionally, the Hessian allow for the estimation of Bayesian evidences (i.e., marginalized likelihoods) for each model using a Laplace approximation, and thus facilitate Bayesian model selection (provided that the posterior is dominated by a single mode).

\subsection{\glom Model Description} \label{sec:model}

\glom enables analyses using a class of GP models, described in \citetalias{Jones2017}, that expresses each of the observable quantities $q_i(t)$ as a linear combination of a single GP, $X(t)$, and its derivatives (Eqs. \ref{eq:class1}-\ref{eq:class3}).
The \glom models are a flexible way to describe complex physical behaviors of multiple observables that are inherently linked together by a latent (i.e., unknown), stochastic function. 

\begin{equation}
 q_0(t) = m_0(t) + a_{00}X(t) + a_{01}\dot{X}(t) + a_{02}\ddot{X}(t) 
%  + a_{04} Z_0(t)
 + \epsilon_{0}(t)
 \label{eq:class1}
\end{equation}

\begin{equation}
 q_1(t) = m_1(t) + a_{10}X(t) + a_{11}\dot{X}(t) + a_{12}\ddot{X}(t)
%  + a_{14} Z_1(t)
 + \epsilon_{1}(t)
 \label{eq:class2}
\end{equation}

\centerline{$\vdots$}

\begin{equation}
 q_l(t) = m_l(t) + a_{l0}X(t) + a_{l1}\dot{X}(t) + a_{l2}\ddot{X}(t)
%  + a_{l4} Z_l(t) 
 + \epsilon_{l}(t)
 \label{eq:class3}
\end{equation}

\noindent 
where 
%$q_i(t)$ are the independent variables meant to be modelled, 
$m_i(t)$ are the mean functions for each $q_i(t)$, $a_{ij}$ are hyperparameters that control the relative amplitude of the GP components, and $X(t)$ is the latent GP that links the outputs, and $\epsilon_{i}$ are measurement uncertainties. 
For the sake of this paper, we will assume that
%that $\epsilon_{i}(t)$ and $\epsilon_{j}(t)$ are uncorrelated with each other. Additionally, we assume 
%observations are made at a discrete set of times and
measurements errors $\epsilon(t)$ are white noise drawn from independent normal distributions with zero mean and known measurement uncertainties. 
Since the measurement errors at different times are uncorrelated, the covariance matrix ($\Sigma_{\epsilon}$) between $\epsilon_i$'s is either a diagonal matrix with the measurement variances ($\sigma_i^2(t_p) = \sigma_{i,p}^2$), or a block diagonal matrix of the measurement covariances at each time ($\Sigma_{\epsilon,p}$).

%
%\begin{equation}
%  \epsilon_{i}(t_p) \sim N(0,\sigma_{i,p}^2)
%  \label{eq:ire}
%\end{equation}
%
%\noindent where $\sigma_{i,p}$ are known measurement uncertainties for ($q_i(t_p)$), the $i$th dependent variable at time $t_p$. 
%

The \glom model requires hyperparameters that will be referred to as $\theta = (\phi, a_{00}, .., a_{l2})$, where $\phi$ is the set of hyperparameters for the chosen kernel function that controls the behavior of $X(t)$ (described in \S\ref{sec:kernel}).
Since the \glom model utilizes $\ddot{X}(t)$, it requires that $X(t)$ be at least twice mean-square differentiable.
This constrains the choice of GP kernels as will be discussed in \S\ref{sec:kernel}.

\subsection{Optimization and Marginalization with the \glom model} \label{sec:optimize}

The GP model that is the most compliant with the data ($\textbf{x}$), and potentially our prior beliefs ($p(\theta)$), is typically found by maximizing the log likelihood of a GP, $\text{ln}(p(\theta|\textbf{x})) = \ell(\theta|\textbf{x})$, or the log unnormalized posterior, $\ell(\theta|\textbf{x}) + \text{ln}(p(\theta))$, when priors have been elicited. 
The log likelihood ($\ell$) for hyperparameters ($\theta$), mean function ($\boldsymbol{\mu}$), and data ($\textbf{x}$) is equivalent to evaluating the probability density function of a multivariate normal distribution with mean, $\boldsymbol{\mu}$, and covariance, $\Sigma$ \citep{Rasmussen2006}.
Thus, the log likelihood is given by 
\begin{equation}
 \ell(\theta|\textbf{x})=-\dfrac{1}{2} \left(N \ \text{log}(2\pi)+\text{log}(|\Sigma|)+(\textbf{x}-\boldsymbol\mu)^T \Sigma^{-1} (\textbf{x}-\boldsymbol\mu) \right),
 \label{eq:gplikelihood}
\end{equation}
\noindent where $N$ is the dimensionality of the normal distribution (in our example application, the number of measurements $N = N_{\text{meas}}=l\times~N_{\text{obs}}$, where $N_{\text{obs}}$ is the number of observation epochs).
% Moving some of this...
In our case, $\Sigma = \Sigma_{GP} + \Sigma_{\epsilon}$ where $\Sigma_{GP}$ is the covariance of the GP components of the \glom model (see Eqs.\ref{eq:J17cov1}-\ref{eq:GPcov}) and $\Sigma_{\epsilon}$ is the covariance of the measurement errors .
$\boldsymbol{\mu}$ is constructed from the mean functions of each $q_i$ such that $\boldsymbol{\mu} = \{m_0(t_0), m_1(t_0), ... , m_l(t_0), m_0(t_1), m_1(t_1), ... , m_l(t_1), ... , m_0(t_{N_{\text{obs}}}), m_1(t_{N_{\text{obs}}}), ... , m_l(t_{N_{\text{obs}}})\}$.

We find that the estimation of the best fit for the \glom model parameters, $\theta$, in the context of our example application (see \S\ref{sec:ex_app}) is vastly more efficient when our fitting algorithms can use the gradient %and Hessian 
of the likelihood
\begin{equation}
 \dfrac{\partial \ell}{\partial\theta_m} = -\dfrac{1}{2}\left(\text{tr}(\Sigma^{-1}\dfrac{\partial\Sigma}{\partial\theta_m}) - (\textbf{x}-\boldsymbol\mu)^T \Sigma^{-1}\dfrac{\partial\Sigma}{\partial\theta_m} \Sigma^{-1}(\textbf{x}-\boldsymbol\mu)\right),
 \label{eq:gradgplikelihood}
\end{equation}
\noindent where $\dfrac{\partial\Sigma}{\partial\theta_m}$ is the partial derivative of the covariance matrix with respect to the relevant hyperparameter. 

Similarly, the optimization process converges an additional factor of 5-10x faster when using methods that incorporate the Hessian information, which can be computed explicitly as
\begin{equation}
 \dfrac{\partial^2 \ell}{\partial\theta_m\partial\theta_n} = -\dfrac{1}{2}\left(\text{tr}\left( \Sigma^{-1} \left( \dfrac{\partial^2\Sigma}{\partial\theta_m\theta_n} - \dfrac{\partial\Sigma}{\partial\theta_n} \Sigma^{-1} \dfrac{\partial\Sigma}{\partial\theta_m}\right)\right) - (\textbf{x}-\boldsymbol\mu)^T \Sigma^{-1} \left(\dfrac{\partial^2\Sigma}{\partial\theta_m\theta_n} - \dfrac{\partial\Sigma}{\partial\theta_m} \Sigma^{-1} \dfrac{\partial\Sigma}{\partial\theta_n} - \dfrac{\partial\Sigma}{\partial\theta_n} \Sigma^{-1} \dfrac{\partial\Sigma}{\partial\theta_m}\right) \Sigma^{-1} (\textbf{x}-\boldsymbol\mu)\right)
 \label{eq:hessgplikelihood}
\end{equation}
Access to the analytical Hessian also allows us to distinguish between local maxima and saddle points in the likelihood or unnormalized posterior space.
Avoiding getting trapped in saddle points is a common problem for multidimensional non-linear optimization problems. 
As a further benefit, we can use the Hessian matrix to estimate the Bayesian evidence for a model using the Laplace approximation % (\ref{eq:laplace}).
\begin{equation}
 \text{ln}\left(\int p(\theta|\textbf{x}) \ p(\theta) \ d\theta \right) \approx \ell(\theta_0|\textbf{x}) + \text{ln}(p(\theta_0)) + \dfrac{1}{2} (N_{\theta} \ \text{ln}(2 \pi) - \text{ln}(\left|-H(\theta_0)\right|))
 \label{eq:laplace}
\end{equation}
\noindent where $\theta_0 = \mathrm{argmax}_\theta \, \ell(\theta|\textbf{x}) + \text{ln}(p(\theta))$ is a maximum posterior estimate for $\theta$, $N_{\theta}$ is the number of hyperparameters, and $H(\theta_0)$ is the Hessian matrix for the unnormalized posterior evaluated at $\theta_0$. 
This approximates the posterior integrand as a Gaussian centered at $\theta_0$ with curvature $-H$. 
The Laplace approximation typically provides a good estimate of the posterior assuming that the posterior mass is dominated by the mode at $\theta_0$, as is typical for our example application once the data identify the proper orbital period for a planet.
The approximation of evidences enables the use of Bayesian model comparison to quantify how well the model fits the data or to compare the quality of models with different kernel functions or numbers of planets.

\section{\glom implementation}
\label{sec:implentation}

\subsection{\glomjl package}
We have implemented the general-purpose \glom model in the open-source package \glomjlns\footnote{\href{https://github.com/christiangil/GPLinearODEMaker.jl}{https://github.com/christiangil/GPLinearODEMaker.jl}} \citep{GilbertsonZenodo}. 
% (doi:\href{https://doi.org/10.5281/zenodo.4144106}{10.5281/zenodo.4144106} \citep{GilbertsonZenodo})
The package is implemented in {\tt Julia} \citep{Bezanson2017}, a modern, efficient language designed for high-performance scientific computing. 
Julia combines the benefits of a high-level language, the performance of {\tt C} or {\tt FORTRAN}, a rich dynamic type system, and a modern package management system. 
\glomjl is thread safe and can be used in either a shared-memory or distributed-memory environment (e.g., when evaluating models with thousands of possible planetary configurations). 
A simple code demonstrating how to use \glomjl is shown in \S\ref{sec:samplecode}. 
Further details and examples are provided in the package documentation. 

\subsection{Constructing the covariance matrix} \label{sec:covariance}

The \glom model is nontrivial to construct and perform inference with because of the correlations between each of the GP outputs.
Linear operations on independent GPs always yield a valid new GP,
since element-wise linear operations of two positive definite matrices always result in another positive definite matrix. 
Thus, taking linear combinations of GPs is relatively straightforward if they do not share hyperparameters, as one can element-wise add or multiply the two covariance matrices together to construct the new GP's covariance matrix.
For combining dependent GPs, one needs to take into account cross terms when constructing the resultant GP's covariance matrix.
The covariance between a GP, $X(t)$, and itself at times $t$ and $t'$ is $\langle X(t),X(t') \rangle = k(t,t')$. 
The covariance between any two dependent variables in the \glom models at any times, $\langle q_i(t), q_j(t') \rangle = k_{ij}^{J17}(t,t')$, can be found with the following

\begin{equation}
  \label{eq:J17cov1}
  k_{ij}^{J17}(t,t') = \sum_{k_1=0}^2\sum_{k_2=0}^2 a_{ik_1} a_{jk_2} \langle\dfrac{d^{k_1}}{dt^{k_1}}X(t), \dfrac{d^{k_2}}{dt'^{k_2}}X(t') \rangle
%\end{equation}
%
%\noindent which can be readily simplified to
% \noindent The differential operators can be taken out of the inner covariance calculations, which allows the following simplification
%
%\begin{equation}
% k_{ij}^{J17}(t,t') 
 = \sum_{k_1=0}^2\sum_{k_2=0}^2 a_{ik_1} a_{jk_2} \dfrac{d^{k_1}}{dt^{k_1}} \dfrac{d^{k_2}}{dt'^{k_2}} k(t,t').
 %\label{eq:J17cov2}
\end{equation}

\noindent For kernels which are a function of absolute separation in time

\begin{equation}
 k_{ij}^{J17}(|t-t'|) = \sum_{k_1=0}^2\sum_{k_2=0}^2 (-1)^{k_2} a_{ik_1} a_{jk_2} \dfrac{d^{k_1+k_2}}{dt^{k_1+k_2}} k(|t-t'|)
 \label{eq:J17cov}
\end{equation}

The covariance matrix is populated by evaluating $k_{ij}^{J17}(t,t')$ at every pair of outputs for all dependent variables.
More specifically, we construct the total covariance, $\Sigma_{GP}$, such that the smaller covariance matrices for each pair of times, $\Sigma_{GP}(t,t')$, are kept together in blocks.
This improves numerical stability for later operations by keeping the dominant terms of the covariance (when $|t-t'|$ is small) close to the diagonal of the matrix (see below).

% \begin{align*}
\begin{equation}
\Sigma_{GP}(t,t') = 
\begin{pmatrix}
k_{00}^{J17}(t,t') & k_{01}^{J17}(t,t') & k_{02}^{J17}(t,t') \\
k_{10}^{J17}(t,t') & k_{11}^{J17}(t,t') & k_{12}^{J17}(t,t') \\
k_{20}^{J17}(t,t') & k_{21}^{J17}(t,t') & k_{22}^{J17}(t,t') \\
\end{pmatrix}
\label{eq:GPcov_parts}
\end{equation}
% \end{align*}

% \begin{align*}
\begin{equation}
\Sigma_{GP} = 
\begin{pmatrix}
\Sigma_{GP}(t_1,t_1) & \Sigma_{GP}(t_1,t_2) & \Sigma_{GP}(t_1,t_3) & \Sigma_{GP}(t_1,t_4) & \ldots\\
\Sigma_{GP}(t_2,t_1) & \Sigma_{GP}(t_2,t_2) & \Sigma_{GP}(t_2,t_3) & \Sigma_{GP}(t_2,t_4) & \\
\Sigma_{GP}(t_3,t_1) & \Sigma_{GP}(t_3,t_2) & \Sigma_{GP}(t_3,t_3) & \Sigma_{GP}(t_3,t_4) & \\
\Sigma_{GP}(t_4,t_1) & \Sigma_{GP}(t_4,t_2) & \Sigma_{GP}(t_4,t_3) & \Sigma_{GP}(t_4,t_4) & \\
\vdots & & & & \ddots\\
\end{pmatrix}
\label{eq:GPcov}
\end{equation}
% \end{align*}

\subsection{Symbolic computation of kernel derivatives}\label{ssec:comp}

% This project is computationally expensive and memory intensive. 
% To combat this, the GP fitting methodology has been entirely implemented in $\mathtt{Julia}$, a new, extremely efficient, and parallelizable coding language. 
% % Our code for this work automatically recognizes when the covariance matrix constructor is passed pairs of observation times that are identical or equally spaced and exploits these properties to eliminate redundant calculations.
% The computational time is largely limited by the Cholesky factorization which is $\mathcal{O}(n^3)$. 

Since we include $\ddot{X}(t)$ in the \glom model, $X(t)$ needs to be twice differentiable. Thus, we need to compute fourth-order time derivatives of the kernel function and second-order derivatives of the kernel function with respect to each non-coefficient hyperparameter (i.e. $\phi$, not $(a_{00} ... a_{l2})$). 
Computing such derivatives analytically for nontrivial kernels is tedious and error prone. 
While derivatives can be computed via autodifferentiation, this results in a significant performance penalty. 
Therefore, \glom symbolically computes and stores the functional forms of all the required derivatives of the basic kernel function provided utilizing SymEngine, a computer algebra system \citep{Meurer2017}.
The compiler optimizes the resulting code to provide efficient evaluation of all required kernel derivatives. 
As an example, consider the code for computing and storing the functional forms of all the required derivatives for the SE kernel.
\begin{minted}{julia}
 function se_kernel_base(lam::Number, del::Number) 
  return exp(-del ^ 2 / (2 * lam ^ 2))
 end
 @vars delta lambda
 GLOM.kernel_coder(se_kernel_base(lambda, delta), "se")
\end{minted}
\noindent The first three lines define our kernel function, the fourth line tells the computer algebra system which variables to treat as symbolic, and the fifth line calls a function to compute and store all of the required derivatives of $\mathtt{se\_kernel\_base}$ and save the resulting source code as $\mathtt{se\_kernel.jl}$.
More complicated kernels can be easily constructed by multiplying or adding any amount of the basic kernel functions together.
For example, the quasiperiodic (QP) kernel \citep[commonly used in analyzing exoplanet survey data]{Aigrain2012} is not explicitly defined in the code, but it is created by combining the necessary squared-exponential components.
\begin{minted}{julia}
 @vars delta delta_P lambda_SE amp_P
 GLOM.kernel_coder(se_kernel_base(lambda_SE, delta) * se_kernel_base(1 / amp_P, delta_P), 
  "qp"; periodic_var="delta_P")
\end{minted}
\noindent where periodic\_var=``delta\_P'' indicates to kernel\_coder that the rotation described in \S\ref{sec:qpk} should be performed on delta\_P%
, and amp\_P $= 1/\lambda_P$ as described in \S\ref{sec:kernel}. 
\glom calculates the needed derivatives analytically once and caches them to efficiently construct the covariance matrix (described in \S\ref{sec:covariance}) and to calculate likelihood gradients and Hessians (described in \S\ref{sec:optimize}).

\section{Example application of \glom to EPRV exoplanet surveys} \label{sec:ex_app}
The analysis of EPRV exoplanet survey was the original motivation the development of the \glom package.
In this section, we validate the \glom model for the analysis of spectroscopic time series that include both planetary Doppler shifts and intrinsic stellar variability. 
First, we provide an overview of the analysis of Doppler exoplanet survey data and the challenges posed by stellar variability in \S\ref{sec:ex_context}. 
%We describe a modern approach to jointly modeling planetary perturbations and stellar activity in \S\ref{sec:joint_modeling}.
%
%
Then, we describe the specific input data used to demonstrate the \glom model in \S\ref{sec:InputData} and \S\ref{ssec:indicators}. 
Next, we describe one possible specialization of the \glom model for Doppler planet surveys (\S\ref{sec:SpecificModel}) and a few potential kernel functions for Doppler exoplanet surveys (\S\ref{sec:kernel}). 
We show example results in \S\ref{ssec:firstresult} and demonstrate the power of \glomjl by comparing results for multiple choices of the covariance kernel in \S\ref{ssec:results}. 

\subsection{Context for Example Application}
\label{sec:ex_context}
As EPRV surveys improve, their ability to detect small planets is increasingly limited by the challenge of separating true Doppler shifts from spurious apparent Doppler shifts caused by intrinsic stellar variability \citep{Fischer2016}. 
Here we review the\deleted{ incoming data,} traditional data reduction strategies, the complications due to stellar activity, and recent progress in addressing this challenge.

\subsubsection{Analysis of Doppler Exoplanet Surveys in the Absence of Stellar Variability} \label{sec:analysis_no_variability}
\explain{We removed the out of scope explanations of the extraction of 1-d spectra for Doppler surveys}

\deleted{Given the complexity of stellar spectra and modern EPRV spectrographs, it can be useful to divide the data analysis procedure for exoplanet surveys into four steps. 
First, standard astronomical image processing techniques are applied to images of the focal plane. Then, each order resulting from dispersion by the echelle grating is extracted, resulting in a list of 1-d spectra, each over a relatively small portion of the instrument's total spectral range. These are often archived as ``2-d extracted spectra'' or ``level 1'' data products. 
Second, the 2-d extracted spectra are combined to produce a ``1-d extracted spectra'' (often labeled as a ``level 2 data product''). Depending on the instrument design, there may be overlap or gaps in the wavelengths spanned by neighboring orders. The instrument throughput often varies significantly near the edge of each order, so care must be taken in the relative normalization and weighting of data from different orders.

Third, the extracted spectra (containing $\sim 10^{5-6}$ flux measurements and uncertainties) are reduced to a small number of summary statistics and their uncertainties.
%\added{EPRV spectrographs and their data processing pipelines are built to estimate the Doppler shifts of the stars they are used to survey.} \cjg{I'm not sold on this intro line}
Since the desired Doppler shifts are a small fraction of the width of a single line or pixel, information needs to be aggregated across many lines to measure Doppler shifts with sufficient precision to detect exoplanets.}
\replaced{Therefore, the apparent Doppler shift is typically derived by first computing}{Apparent Doppler shifts from Doppler exoplanet surveys are typically estimated by maximizing} a cross-correlation function (CCF) between the observations and either a high signal-to-noise ``template'' spectrum or a CCF ``mask'' \citep[often taken to be a weighted sum of top-hat functions near the location of thousands of spectral lines; e.g.,][]{Zechmeister2018,Pepe2002}. %\TODO: or Gaussians or is that a new development/not widely used?
\deleted{The reported radial velocity is based on the Doppler shift to the template that maximizes the CCF for a given observation.}
In an idealized scenario (e.g., perfect template, the intrinsic stellar spectrum is fixed, no contamination from telluric features or scattered light, instrumental line spread function is fixed), the traditional \added{CCF} approach corresponds to finding the Doppler shift for the template that maximizes the standard $\chi^2$ statistic for comparing the observations and Doppler-shifted template.
However, if the intrinsic stellar spectrum is changing, then the theoretical basis for the CCF method breaks down, motivating the joint analysis of apparent Doppler shifts and stellar variability (see \S\ref{sec:joint_modeling}).

%Dimension reduction techniques (e.g., PCA) can also be used to construct high-information stellar activity indicators from the data instead of specifying types of stellar activity in advance \citep{Davis2017}. 
%
%Additionally, bisector velocity spans, line widths, and chromospheric activity indices have also been used \citep{Rajpaul2015}.
%\cite{Jones2017} developed the Doppler-constrained Principal Components Analysis (DPCA) method for analyzing spectroscopic time series. It reduces each spectrum into an apparent radial velocity and multiple DPCA ``scores'' that serve as stellar variability indicators. 
%DPCA aggregates information across spectral lines and also ensures that each stellar activity indicator is orthogonal to the apparent Doppler shift. 

\replaced{Finally}{To characterize exoplanets}, the time series of measured radial velocities (and their associated uncertainties) are modeled as a combination of true planetary signals, uncorrelated measurement noise and, potentially, a contribution due to stellar activity. 
\deleted{When there is an existing measurement for the orbital period of a planet or planet candidate (e.g., based on transit observations), then one can use standard techniques %\citep[e.g., MCMC]{Ford2005,Ford2006} 
to characterize the posterior distribution for planetary parameters in the vicinity of the presumed orbital period. }
When searching for new planets, the analysis is typically broken up into a global search stage (e.g., a brute force search over orbital period using a finely spaced grid, often assuming a circular orbit) followed by a local exploration stage at each potential orbital period (e.g., maximizing likelihood or posterior for the fixed given period, or the likelihood marginalized over all parameters conditioned on the presumed orbital period). 
For systems with multiple planets, one typically iterates the global search stage to find approximate orbital periods for all detectable planets, again using traditional methods to characterize the posterior in the vicinity of a given set of orbital periods. 

\subsubsection{Challenge of Stellar Variability}\label{sec:challenge}

The variability of stellar spectra caused by starspots or pulsations can mimic and obscure planetary RV signals. 
For example, when stellar magnetic fields interact with the envelope of convective gas at the stellar surface (the photosphere), groups of small, dark, and relatively cool starspots, often surrounded by brighter regions called faculae, can form. 
When starspots are on the side of the star that is moving toward the observer, the amount of blue-shifted light is decreased, causing a distortion in the spectral line shapes. 
%At the spectral resolutions used by current RV spectrographs, 
The distortion causes the star to look more red shifted overall. 
As the star rotates, the spot obscures less blue shifted light and more red shifted light, mimicking a periodic planetary Doppler shift. 
Additionally, the magnetic fields associated with starspots also inhibit the strength of the convective blueshift.
Stellar rotation periods can be similar to plausible orbital periods for exoplanets, which can make it especially difficult to separate signals from spots and faculae and planets \citep{Dumusque2012, Rajpaul2015}. 
Additionally, other sources of stellar variability (e.g. granulation, pulsations, convective motions, etc.) can exacerbate the problem \citep{Lovis2010}. 
Contamination from these processes make it difficult to reliably detect low mass planets or planets orbiting magnetically active stars.

\subsubsection{Distinguishing Planetary Perturbations \& Stellar Variability}
\label{sec:joint_modeling}

In principle, subtle differences between how stellar variability and planets affect stellar spectra can be used to distinguish them.
For example, a Doppler shift caused by a (stably orbiting) planet is a strictly periodic signal and affects the entire spectrum in a uniform manner, whereas signals from stellar variability are fundamentally transient and wavelength dependent.
\added{These effects are usually investigated by analyzing summary statistics generated from the spectra and CCF}
\explain{The following was moved here from \S\ref{sec:analysis_no_variability}}
%At present, there is no direct way to measure how the current activity state of stars is affecting their spectra. 
%However, several proxies for stellar activity exist. 
\deleted{In addition to the apparent radial velocity measurements, }Common summary statistics include the measurements of specific strong spectral lines (e.g., Ca II H \& K, H-$\alpha$) and properties of the cross-correlation function between the spectrum and a mask or template spectrum \citep[e.g., the full width half maximum (FWHM), bisector span or bisector slope;][]{Pepe2002}.

More recently, data-driven models have been proposed as a path to developing spectroscopic indicators that are more directly linked to spurious Doppler signals induced by stellar variability. 
These have the advantage of aggregating information from multiple spectra lines, while also allowing for data to determine how the information from different lines should be combined and weighted.
\citet{Davis2017} demonstrated that principal components analysis (PCA) serves as an efficient form of dimensional reduction, at least for solar spectra simulated by SOAP 2.0 \citep{Dumusque2014}, where only 1-3 basis vectors are needed given the spectral resolution and signal-to-noise anticipated for next-generation Doppler planet surveys. 
\citetalias{Jones2017} proposed a Doppler-constrained PCA (DPCA) method that explicitly incorporates knowledge of how a true Doppler \added{shift} affects the spectra. 
A set of spectra are summarized as a Doppler-shifted mean spectrum plus a linear sum of basis vectors (principal components).
Critically, the algorithm ensures that the basis vectors are orthogonal to the perturbations caused by a planetary Doppler shift, as well as each other and so capture different information.
\explain{The preceding was moved here from \S\ref{sec:analysis_no_variability}}

\deleted{If the changes in the stellar spectra with time are due purely to the Doppler effect as the star orbits around the system center of mass, then one would expect no correlation between stellar activity indicators (only varying due to measurement noise) and the orbital phase. 
On the other hand, if intrinsic stellar variability were causing the spectra to change in a way that appears to mimic kinematic Doppler shifts, then there is a significant chance that measurements of stellar activity indicators correlate with the putative orbital phase. 
Therefore, astronomers typically analyze the available stellar activity indicators as well as the measured Doppler shifts.
If both show a similar periodicity, then extreme caution is warranted before claiming to have detected an exoplanet. 
In short, the state-of-the-art consists of first assuming that stellar variability is not significantly affecting the observed spectra, and then checking if there's observational evidence to refute this assumption.}
\added{Traditionally, when analyzing measured RVs, astronomers often initially assume that stellar variability is not significantly affecting the observed spectra and then check for similar periodicities in the apparent RVs and summary statistics.
If found, then extreme caution is warranted before claiming to have detected an exoplanet.}
\replaced{The traditional approach described above}{This approach} has powered the detection of hundreds of exoplanets. 
At the same time, new approaches are needed to maximize the scientific return of current and upcoming EPRV exoplanet surveys and to enable the next-generation of EPRV surveys to detect potentially Earth-like planets around sun-like stars. 
In order to achieve these goals, astronomers will need to: (1) develop methods to measure the Doppler shift robustly, (2) identify and measure stellar variability indicators that are correlated with the contamination to apparent Doppler shifts due to stellar variability, and (3) apply a statistical model for jointly modeling the measured Doppler shift and stellar variability indicators. 
The \glom model addresses the third need using multivariate GPs to jointly model apparent RV signals and stellar activity indicators. 
In the example below, we use the approach of \citetalias{Jones2017} to measure Doppler shifts and stellar activity indicators from the input spectra described in \S\ref{sec:InputData}. 
However, the \glomjl package is designed to be general, so that astronomers can easily substitute their preferred method for measuring Doppler shifts and preferred stellar variability indicators. 
We anticipate that providing a common statistical framework and tools will help advance the state of EPRV data analysis by facilitating apples-to-apples comparisons of different EPRV analysis methods and potential stellar variability indicators.

\subsection{Input Data}
\label{sec:InputData}
%
%In this work, we further the study of the effectiveness of the \glom models by testing their ability to detect planets in the presence of a realistic distribution of evolving starspots.
%To do this, we will use the empirically-motivated time series of simulated solar spectra described in \citet{Gilbertson2020}.
To test \glomns, we model radial velocities and activity indicators calculated using simulated solar spectra described in \citet{Gilbertson2020}, hereafter \citetalias{Gilbertson2020}.
\citetalias{Gilbertson2020}, used SOAP 2.0 \citep{Dumusque2014} to construct several decades of empirically-informed solar spectra simulations that follow activity levels described in the literature.
The spectra in this data set are built using physically motivated distributions for spot areas, decay rates, and latitudes as well as differential rotation rates depending on stellar spot latitudes.
The RV measurements from this data set have a RMS deviation from 0 of $\sim 0.82$ m/s. 
We construct simulated data sets by randomly selecting a subset of 100 observations from one year of simulated observations in the \citetalias{Gilbertson2020} data set. 
% For data sets with a planet, we set the orbital period to be $P=$sqrt(72)$\approx$8.485 days (an irrational period to avoid sampling aliases).
%

\subsection{Measuring Apparent Doppler Shift \& Stellar Variability Indicators} \label{ssec:indicators}

In our example application, we use apparent RVs and indicators that were calculated by applying DPCA \citepalias{Jones2017} on the solar-like simulated spectra from \citetalias{Gilbertson2020}.
For our analysis, we compute basis vectors using the noiseless spectra from \citetalias{Gilbertson2020}. 
Then, we simulated practical measurements by adding photon noise to each pixel in the spectra with an average signal to noise ratio (S/N) of 100 (per resolution element).
Finally, for each observation, we compute the apparent RV, weights for each basis function (principal component ``scores''), and their associated measurement uncertainties (including their covariances). % based on generating hundreds of Monte Carlo realizations of the observed spectra with different levels \cjg{might be confusing, should we say something like draws?} of photon noise. 
Given the high spectral resolution in the \citetalias{Gilbertson2020} simulations and the assumed S/N, each observation results in a single measurement RV precision of $\approx$ 24 cm/s, comparable to the design specifications of state-of-the-art spectrographs. 
Following \citetalias{Jones2017}, we use 2 (Doppler-constrained) principal components, DPCA$_1$ and DPCA$_2$, in addition to the apparent RV, as this is sufficient to accurately reconstruct the simulated solar spectra given the assumed resolution and S/N.  %(see figure \ref{fig:pcavariance})
%
%\ref{ssec:soap}
%data set to calculate our scores.
%It is beyond the scope of this work to explore the best way to find the basis vectors in the presence of realistic observational conditions.
%After training, but before calculating scores, 
%We compute the apparent radial velocity 
% Pointwise errors for the calculated components at each time are calculated based on bootstrap resampling of the Doppler-constrained PCA scores with the same noise level.

\subsection{Example Model}
\label{sec:SpecificModel}

We analyze the data sets simulated above using the multivariate GP model recommended by \citetalias{Jones2017} based on its performance in detecting planets in the presence of a single non-evolving spot. 
%Hereafter, we will only be referring to the \glom model that they found to perform the most favorably for detecting planets
The recommended model (reproduced in Eqs. \ref{eq:model1}-\ref{eq:model3}) is a specific case of the \glom model, where several of the $a_{ij}$ coefficients have been fixed at zero.

\begin{equation}
 \widehat{\text{RV}}(t) = m_0(t) + a_{00}X(t) + a_{01}\dot{X}(t)
%  + a_{04} Z_0(t)
  + \epsilon_{0}
 \label{eq:model1}
\end{equation}

\begin{equation}
 \widehat{\text{DPCA}}_1(t) = m_1(t) + a_{10}X(t) + a_{12}\ddot{X}(t)
%  + a_{14} Z_1(t)
  + \epsilon_{1}
 \label{eq:model2}
\end{equation}

\begin{equation}
 \widehat{\text{DPCA}}_2(t) = m_2(t) + a_{21}\dot{X}(t)
%  + a_{l4} Z_l(t) 
  + \epsilon_{2}
 \label{eq:model3}
\end{equation}

%\subsection{Mean functions}
\label{sec:means}

The mean functions $m_1(t)$ and $m_2(t)$ are set to zero, as we expect all of the behavior to be modelled by the GP components. 
When one (or more) planets are present, $m_0(t)$ is set to the radial velocity perturbation predicted given the planet masses and orbits. 
In this example, we consider a single planet on a Keplerian orbit. 
Thus, $m_0(t)$ is parameterized by $K$ (the radial velocity amplitude), $P$ (the orbital period), $e$ (the orbital eccentricity), $\omega$ (the periastron direction), and $M_0$ (the mean anomaly at epoch). 
When considering models with no planets, $m_0(t)$ is also set to zero.
%apparent radial velocity $\widehat\text{RV}}(t)$ 
%We wish to test the activity distinguishing capabilities of this \glom model.
%
%\ebf{What do we do about epsilons? Are we assuming the components are independent? Drawing from a covariance matrix that accounts for any correlations? }
%\cjg{I think we actually use a bootstrap estimated 3x3 covariance matrix at each observation. thats what drove us to rearranging the covariance matrix to put things along the diagonal}
We assume $\epsilon(t_p) \sim N(0,\Sigma_{\epsilon})$, where the measurement covariance of the apparent RV and spectroscopic indicators at each epoch is treated as known.

\subsection{GP Kernel functions} \label{sec:kernel}

The stochastic behavior of any GP, and the \glom model in particular, is controlled by the choice of kernel function. 
Each kernel function requires one or more parameters which are often called hyperparameters and we will refer to them generally as $\phi$.
Below we summarize key properties of several kernels that are implemented in \glom and are of interest for our example application to stellar variability. 
The quality of the GP model fits to our simulated observations using different choices for the covariance kernel will be compared in \S\ref{ssec:results}. 

\subsubsection{White-noise Kernel}
The simplest kernel choice is a white noise kernel, %(\ref{eq:whitekf}). 
\begin{equation}
 k_{white}(t,t') = \delta(t-t')
 \label{eq:whitekf}
\end{equation}
This kernel assumes that any two draws from the GP at different values of the independent variable are uncorrelated, so realizations from this GP are extremely rough. Using this kernel for each observable in the context of a stellar activity model would be equivalent to modeling the effects from stellar variability as stellar ``jitter'' \citep[e.g.,][]{Ford2006}.
%However, the resulting GPs are not usable within the context of the \glom model. 
%\cjg{I mean, they are \textit{technically} usable. The jitter runs were actually a GLOM model with only a00 and a delta function. It's just that if that were the model you wanted to use, you shouldn't waste your time doing it with all of this extra complex machinery, hence the original, now-commented line}
In this case, the covariance matrix would be block diagonal (blocks since measurement uncertainties can cause the apparent velocity and stellar activity indicators at each epoch to be correlated). Therefore, the matrix factorization could be completed very efficiently, and the \glomjl package would be unnecessarily complex for such a simple model.
%, although more efficient non-GPR fitting methods should be used when modeling the noise this way. 

\subsubsection{SE Kernel}
One kernel function that has been used to model stellar variability (specifically in stellar photometry, see \citet{Liu2018, Littlefair2017, Aigrain2012, Aigrain2015}) is the squared exponential (SE) kernel (\ref{eq:sekf}). %
%The SE kernel is especially suited toward fitting smooth functions and is defined in the following way
%
\begin{equation}
 k_{SE}(t,t') = \text{exp}\left(-\dfrac{(t-t')^2}{2 \lambda_{SE}^2}\right)
 \label{eq:sekf}
\end{equation}
\noindent where $\lambda_{SE}$ is the time-scale of local correlations\footnote{A $\sigma^2$ prefactor which controls the amplitude of the covariance is often included in kernel functions, but is unnecessary within the context of \glom models, as $\sigma^2$ would be degenerate with re-scaling all of the the $a_{ij}$'s.}.
%Parameters in the kernel function, like $\lambda_{SE}$, are often called hyperparameters and we will refer to them as $\phi$.
% The set of hyperparameters for a given kernel will hereafter be generally referred to as $\phi$.
%
The SE kernel is a common choice for the kernel function in GP regression problems, as it can approximate any continuous function on any subset of the input space and draws from a GP with a SE kernel yield samples that are infinitely differentiable \citep{Duvenaud2014}. 
The combination of a single characteristic timescale and extreme smoothness can cause the SE kernel to struggle for modeling processes where there are sharp changes, e.g., when an active region appears on the surface or rotates into view.

%\subsubsection{Exponential Kernel}
%
%The simplest kernel that still allows for covariances between data points is the exponential kernel (\ref{eq:expkf}), also sometimes called an Ornstein-Uhlenbeck process \citep{Rasmussen2006}.
%
%\begin{equation}
% k_{Exp}(t,t') = \text{exp}\left(-\dfrac{|t-t'|}{\lambda_{Exp}}\right)
% \label{eq:expkf}
%\end{equation}
%

\subsubsection{Mat\'ern Kernel}
The Mat\'ern family of kernels is more suited toward modeling rougher behavior.
%The Mat\'ern kernels produce behavior between the infinite smoothness of the SE kernel and the roughness of an exponential kernel.
Of particular interest to this work is the Mat\'ern $^5/_2$ (M$^5/_2$) kernel, %(\ref{eq:m52kf})
\begin{equation}
 k_{M^5/_2}(t,t') = \left(1 + \Delta t + \dfrac{\Delta t^2}{3}\right) \ e^{-\Delta t}
 \label{eq:m52kf}
\end{equation}
\noindent where $\Delta t = \sqrt{5} \ |t-t'| / \lambda_{M^5/_2}$ and $\lambda_{M^5/_2}$ is the timescale of local variations.
It is the lowest order Mat\'ern kernel that can be calculated quickly and yields a GP that is at least twice mean-square differentiable, as required for use with the full \glom model.% because it utilizes $\ddot{X}(t)$.

\subsubsection{QP Kernel}
\label{sec:qpk}
Another common kernel used for modeling stellar variability in both photometry \citep{Grunblatt2015, Aigrain2012, Angus2018} and radial velocities \citep{Jones2017, Rajpaul2015, Haywood2014, Nava2020} is the quasiperiodic (QP) kernel function %(\ref{eq:qpkf})
\begin{equation}
 k_{QP}(t,t') = \text{exp}\left(-\dfrac{\text{sin}^2(\pi(t-t')/\tau_P)}{2 \lambda_P^2}-\dfrac{(t-t')^2}{2 \lambda_{SE}^2}\right).
 \label{eq:qpkf}
\end{equation}
\noindent where $\lambda_{SE}$ is the time-scale of local correlations,
$\tau_P$ is the time-scale of periodic correlations, and
$\lambda_P$ describes the relative importance of the periodic and local correlations.  
The QP kernel is used to fit functions that are locally periodic, but do not maintain a constant amplitude or coherent phase.
A QP kernel function is often selected because it can model rotationally modulated variability from long-lived active regions (e.g., spots, faculae) and the varying signal from the temporal evolution of the active region. 

%It should be noted that the QP kernel is the result of multiplying the SE kernel with the periodic kernel, shown in Eqs. \ref{eq:sekf} and \ref{eq:pkf}. The periodic kernel, in turn, is a rotation of the SE kernel, replacing $(t-t')^2$ with $\text{sin}^2(\pi(t-t')/\tau_P)$. 

%\begin{equation}
% k_P(t,t') = \text{exp}\left(-\dfrac{\text{sin}^2(\pi(t-t')/\tau_P)}{2 %\lambda_P^2}\right)
% \label{eq:pkf}
%\end{equation}

While the QP kernel is physically motivated \citep{Aigrain2012}, in practice, it results in extremely smooth behaviour of $X(t)$. It is possible that the stellar variability (or other processes that can be modelled with the \glom model) might be more amenable to modeling from a ``rougher'' kernel function, such as M$^5/_2$. We will explore this possibility in \S\ref{ssec:results}.

% \subsection{Example code?}
% \ebf{Would you want to paste in ~3-10 lines of code showing how to use GLOM?}
% \cjg{would this replace the proposed julia package section? or is this a separate thing}

% \ebf{I'm thinkgin the "why julia" would go somewhere in/near the implementation section. But this would be just showing people how to use it}
% \cjg{gotcha}

\subsection{Example Results} \label{ssec:firstresult}

\begin{figure}
\centering
\includegraphics[width=18cm]{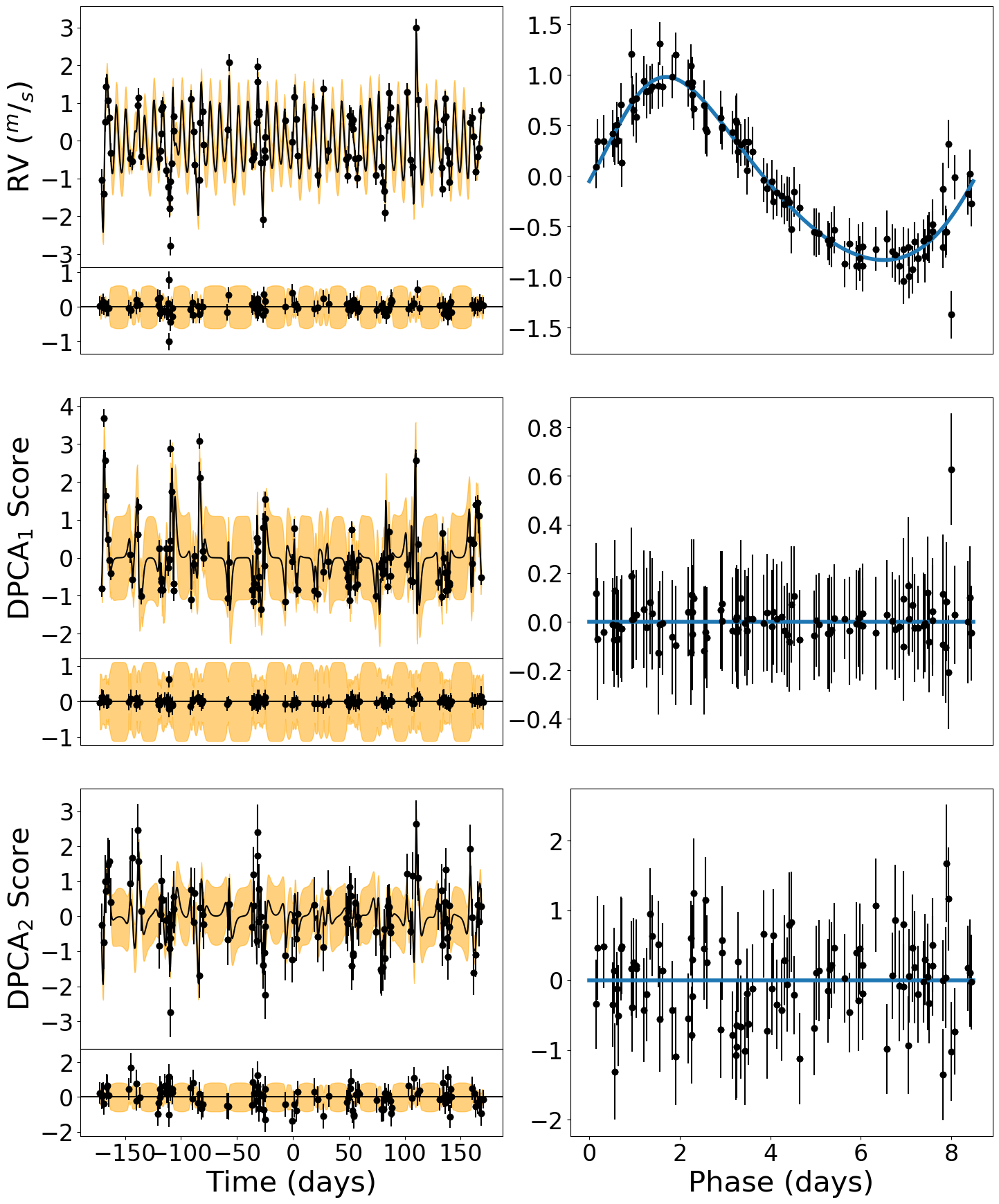}
\caption{A joint fit of a \glom model (Eqs. \ref{eq:model1}-\ref{eq:model3}) with a Keplerian RV mean function. % and the XXX covariance kernel.
%TODO which kernel function? 
The data come from one draw of 100 observation times from the \citetalias{Gilbertson2020} data set of solar-like stellar variability. 
A Keplerian signal has $P=$sqrt(72)$\approx$8.485 days and $K=$1 m/s has been injected into the apparent radial velocities (top row). 
Left: The upper plot in each panel shows the data with 1-$\sigma$ error bars (points), the mean for the MAP \glom model (black line), and the 1$\sigma$ model uncertainties (orange fill). The lower plot in each panel shows the residuals from the mean of the posterior predictive distribution. 
Right: The phase folded residuals from the mean of the posterior predictive distribution at the planet's orbital period. The RV residuals show a clear planetary signal and the DPCA residuals do not show any periodic behavior.}
\label{fig:samplefit}
\end{figure}

We apply \glom to find the maximum a posteriori (MAP) estimate of the \glom model parameters. Priors for the hyperparameters are described in \S\ref{sec:priors}.
%a noisy version of the \citetalias{Gilbertson2020} data set.
The MAP estimate for the $\theta$ of the \glom model is found using the modified version of Newton's method with a trust region, as implemented in {\tt Julia}'s {\tt Optim.jl} package \citep{Mogensen2018}.
For the sake of simplicity, when fitting a planet, we assume that the period is known {\em a priori}
(This corresponds to the science case of using spectroscopic observations to measure the mass and eccentricity of a planet identified via transit observations).
If one wanted to apply the model to search for a planet with unknown orbital period, then this approach could be combined with conventional period search algorithm (typically a brute force grid search).
A sample joint \glom and Keplerian fit to a representative set of 100 observations within one observing season is shown in Fig. \ref{fig:samplefit}.
Here the injected Keplerian signal has an orbital period of $P=$sqrt(72)$\approx$8.485 days (an irrational period to avoid sampling aliases) and an RV amplitude of 1 m/s.
% TODO: What eccentricity was used? 
%
For a fixed orbital period and $\theta$ (kernel hyperparameters), 
% TODO: Do you mean \phi instead of \theta?
% no because this includes the a coefficients
there are only additional two nonlinear model parameters, $M_0$ and $e$.
Therefore, we use an iterative non-linear fitting algorithm \citep[specifically L-BFGS in {\tt Optim.jl};][]{Mogensen2018}) to identify the best-fit values of $M_0$ and $e$, marginalizing over the linear parameters analytically. 
Iterations alternate between holding $\theta$ 
% TODO: Do you mean \phi instead of \theta?
% no because this includes the a coefficients
and the Keplerian parameters fixed, while optimizing for the remaining parameters.
We verify that the algorithm reliably converges on the MAP parameters. 
This approach is analogous to the scheme for fitting Keplerian orbits described in \citet{Wright2009}, but instead of minimizing $\chi^2$, we maximize the posterior using the GP covariance matrix ($\mathbf{\Sigma}$). 
Our approach accounts for the priors (described in \S\ref{sec:keppriors}) and efficiently models the stellar variability and planetary RV to the apparent RV and spectroscopic activity indicators simultaneously.

% \subsection{Performance?}
% \ebf{Would you want to say anything about how fast it is?}
% \cjg{This might be a sore point as we are still n^3}
% \ebf{That's ok. We could even make it into a plot versus n. Although, let's not let this distract us from getting the rest of the paper in submittable shape.}
% \cjg{agree}

\subsection{Impact of GP Kernel Choice on Planet Characterization} \label{ssec:results}
\explain{We made section 5 into a subsection to improve readability.}

The ability of \glomjl to symbolically compute the necessary derivatives makes it easy to explore the impact of different choices for the GP kernel function and specific \glom model. 
Here, we compare the results of modeling our simulated spectroscopic time series using different GP covariance kernels that correspond to different assumptions about the nature of stellar variability:
%performance of the model from \S\ref{sec:model} using multiple GP covariance kernels:
\begin{enumerate}
  
  \item Quasiperiodic (QP): Model apparent RVs and spectroscopic indicators using \S\ref{sec:SpecificModel} and the QP kernel (for both apparent RVs and spectroscopic indicators)
  
  %\item Model apparent RVs and spectroscopic indicators using \S\ref{sec:SpecificModel} and the squared exponential kernel (for both apparent RVs and spectroscopic indicators)
  
  \item Mat\'ern $^5/_2$ (M$^5/_2$): Model apparent RVs and spectroscopic indicators using \S\ref{sec:SpecificModel} and the M$^5/_2$ kernel (for both apparent RVs and spectroscopic indicators),
  
  \item Jitter: Model only apparent RVs as white noise (``jitter term''), i.e., $\epsilon_0(t_p) \sim N(0,\sigma_{0,p}^2 + \sigma_j^2)$, discarding measurements of $\widehat{\mathrm{DPCA}}_i(t)$ for $i = 1,2$ (Eqns.\ \ref{eq:model2} \& \ref{eq:model3}).
  
  \item No Activity Model (NAM): Model only apparent RVs and assume that stellar variability does not result in any perturbations to the apparent radial velocity, i.e., $\epsilon_{0}(t) \sim N(0,\sigma_{0,p}^2)$, discarding measurements of $\widehat{\mathrm{DPCA}}_i(t)$ for $i = 1,2$ (Eqns.\ \ref{eq:model2} \& \ref{eq:model3}). 
  
\end{enumerate}

\begin{figure}
\centering
\includegraphics[width=18cm]{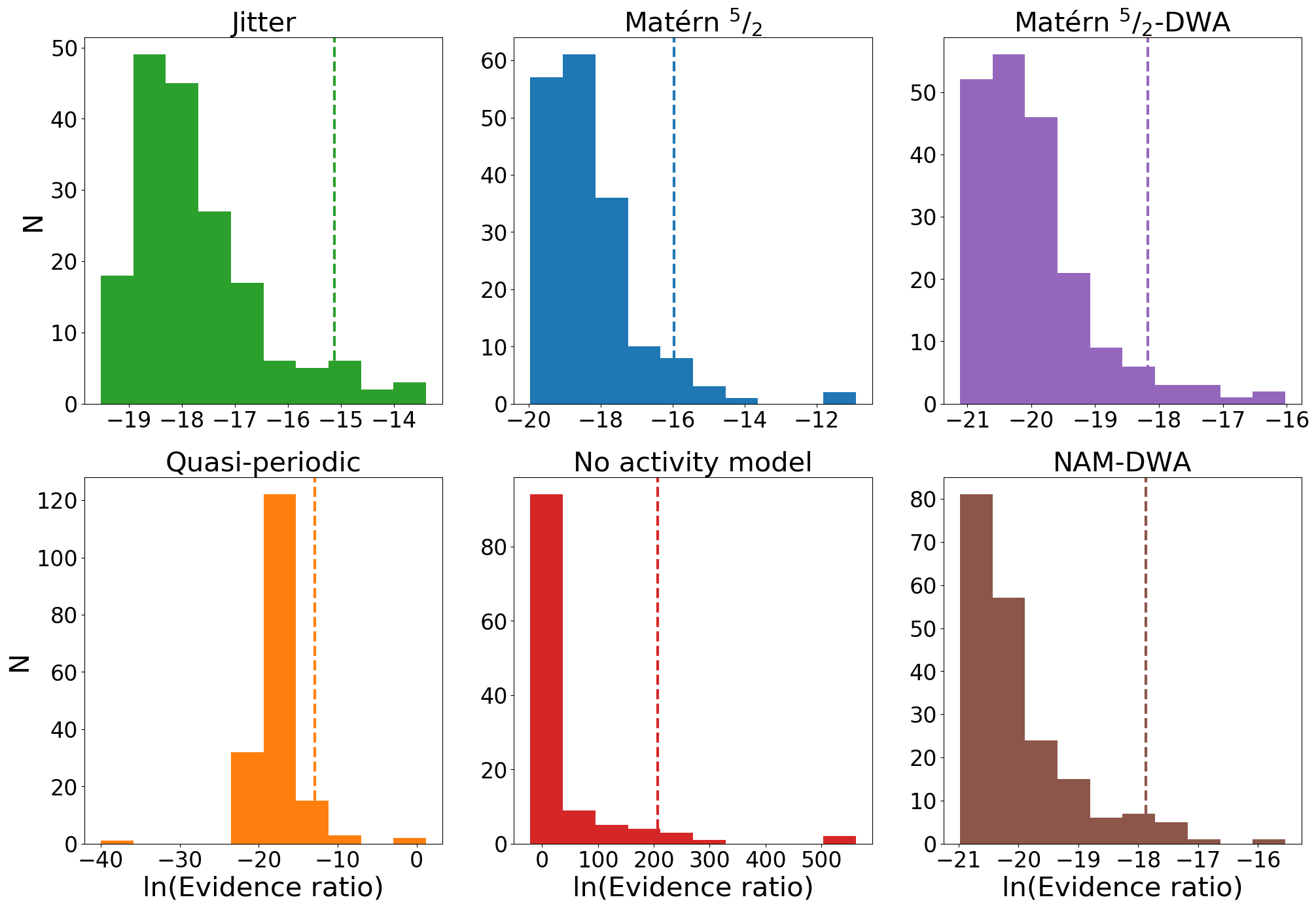}
\caption{The distribution of the estimated log evidence ratio comparing models for stellar variability with or without a planet on a Keplerian orbit. Larger evidence ratios indicate stronger evidence in favor of including a planetary signal. The vertical dashed lines indicate the 95\% quantile. 
Top left: Assuming a white noise (or ``jitter'') model for stellar activity.
Bottom left: Assuming the \citetalias{Jones2017} model with a QP kernel for stellar activity. 
Top middle: Assuming a \glom model with a M$^5/_2$ kernel for stellar activity.
Bottom middle: Applying a no activity model (NAM) on a data set with stellar activity.
%This showcases the worst case scenario for planet detection. 
In each of the above cases, the data analyzed include only simulated stellar activity \citepalias{Gilbertson2020}.
For these cases, a qualitatively larger tail to the right (e.g., upper right, lower left, upper left) implies that the models may have limited power for detecting low-mass planets.
In contrast, a compact distribution suggests that even a low-mass planet could cause the evidence ratio to increase above the detection threshold. 
In the following two cases, the evidence ratio was calculated for data generated without any stellar activity (DWA) and only photon noise.
Top right: Assuming a \glom model with a M$^5/_2$ kernel for stellar activity. 
Bottom right: Applying a no activity model (NAM) on a data set without stellar activity.
Comparing these two shows that there is relatively little cost for incorporating a stellar activity model in the event that a target star has essentially no contamination due to stellar activity. }
\label{fig:erhist}
\end{figure}

For each case, we consider both the no-planet case ($m_0(t)=0$) and the one-planet case ($m_0(t)$ equal to RV for a Keplerian orbit). 
Once we have identified the best-fit parameters for both the 0-planet and 1-planet models, we evaluate the Hessian at the location of the best-fit parameter to estimate the Bayesian evidence (i.e., marginalized likelihood) for each model using the Laplace-approximation \citep{Nelson2020}. % https://doi.org/10.3847/1538-3881%2Fab5190 This just got published?!? I thought it was out in like 2018. Yes, dealing with referee reports can take a long time, especially when people move on to another position.
The evidence ratio quantifies how much better the 1-planet model describes the data than the 0-planet model.
We currently do not impose an imbalanced prior on whether or not a planet exists, thus our evidence ratios are also Bayes factors. 
In each panel of Fig. \ref{fig:erhist}, we show the distribution of log evidence ratios based on 200 simulated data sets. 
The panels differ in the choice of covariance kernel and whether the input data included stellar activity. 
In the bottom center and bottom right panels, we show results based on analyzed data that was generated with no stellar activity.

Often, astronomers desire to establish a detection criterion with a well-characterized false discovery criterion.
A large evidence ratio could arise from a small data set where the 1-planet model results in significantly lower amplitude residuals than the 0-planet model.
Alternatively, a large evidence ratio could arise in a large data set, even if the residuals to the 1-planet model are only slightly smaller than the residuals to the 0-planet model, since there are more measurements. 
Therefore, if one were to choose a critical evidence ratio threshold for detecting a planet that does not depend on the number of observations (or the strength of covariances between observations), then it would be expected that the false discovery rate could be small for a small number of observations, but increase as more observations were taken. 
If one aims to achieve a certain false discovery rate, then one can calibrate the critical evidence ratio by computing the evidence ratio for a large sample of simulated data sets that are comparable to the available observations and do not include any planetary signal. 
The $1-q$ quantile of the evidence ratios for the no-planet simulations sets the minimum evidence ratio required to detect a planet with a false discovery rate of $q$. 
For example, the vertical dashed lines in Fig. \ref{fig:erhist} show the 95\% quantile. 
Of course, critical threshold is specific to survey properties (i.e., number of observations, measurement uncertainties), the properties of the simulated data sets (e.g., true correlations between measurements) and the model assumptions (e.g., assumed GP kernel). 
In this example, the assumption of known period allows for a lower detection threshold than if one had to consider all possible orbital periods.

Comparing the shapes and scales of the evidence ratio distributions in Fig. \ref{fig:erhist} can help us anticipate the ability of each model for detecting planets. 
In the no activity model (NAM, lower middle), stellar activity often leads to signals such that the evidence ratio strongly favors a 1-planet model, since it is extremely unlikely that random measurement noise would replicate stellar activity. 
If the NAM model were applied to an EPRV survey of stars with solar-like stellar variability, then either there would be frequent false alarms or the critical evidence ratio would need to be set so high that it would result in low sensitivity for low-mass planets.
%While not as extreme, we find that the Jitter (top left) and QP models (bottom left) also have qualitatively larger tails to the right. 
Of the models considered, the \glom model with a M$^5/_2$ kernel appears to have the smallest range from the median to the 95\% quantile of evidence ratios. 
This suggests that it has the potential to be a powerful model for detecting low-mass planets in the presence of stellar variability.

While it's clear that using some model for stellar variability is much better than the NAM, one might wonder if there are significant scientific costs to adopting a \glom stellar activity model for a target star that did not have significant stellar variability. 
This can be addressed by studying the distribution of evidence ratios computed for using data sets without activity (DWA). 
Comparing the M$^5/_2$ model (upper center) and M$^5/_2$-DWA model (upper right), we see little difference in the distributions.
Thus, we expect that there will be relatively little loss of sensitivity by applying the \glom model with M$^5/_2$ kernel for stars with limited stellar variability. 
Indeed, these distributions are also quite similar to that of the NAM-DWA model (lower right), i.e., if there were no stellar activity and one ``knew'' that there were no stellar variability. This serves as a limiting case, either an ideal target star with no stellar variability or the output of applying a perfect stellar variability mitigation strategy to a real star. 
The similarity of the evidence ratios for the NAM-DWA and M$^5/_2$ models is encouraging, as it suggests that there may be little scientific cost to using the \glom model with M$^5/_2$ kernel (for the particular survey properties and solar-like variability assumed in these simulations).

%
%
%We then approximate the evidence for the \glom model replacing $m_0(t)$ (which , before, was 0) with a Keplerian component to same data set.
%We leave the problem of searching for Keplerian signals without period constraints for a future work in this series
%, and for now assume that there are strong constraints (say from a prior transit analysis) that allow us to assume a known period of $P=$sqrt(72)$\approx$8.485 days (an irrational period to avoid sampling aliases).

% \cjg{need to talk about model selection in general here i think. or maybe not. not sure}

%Without ground truth knowledge of the phenomena one is trying to model, it is not clear whether the evidence ratio from fitting any given data set is enough to determine whether one model is truly preferred over than another.
%To calibrate our understanding for what evidence ratio we can consider sufficient to say that an additional Keplerian component is preferred, we can observe the distribution of evidence ratios between MAP models with and without a planet component.
%We chose to fit subsets of 100 noisy observations of the RVs and activity indicators.

%for 200 subsets of 100 observations of the noisy \citetalias{Gilbertson2020} data (shown in Fig. \ref{fig:erhist}). 
%These subsets of 100 observations were selected by randomly taking the RVs and year of the \citetalias{Gilbertson2020} data from out of the 730 observations

\begin{figure}
\centering
\includegraphics[width=18cm]{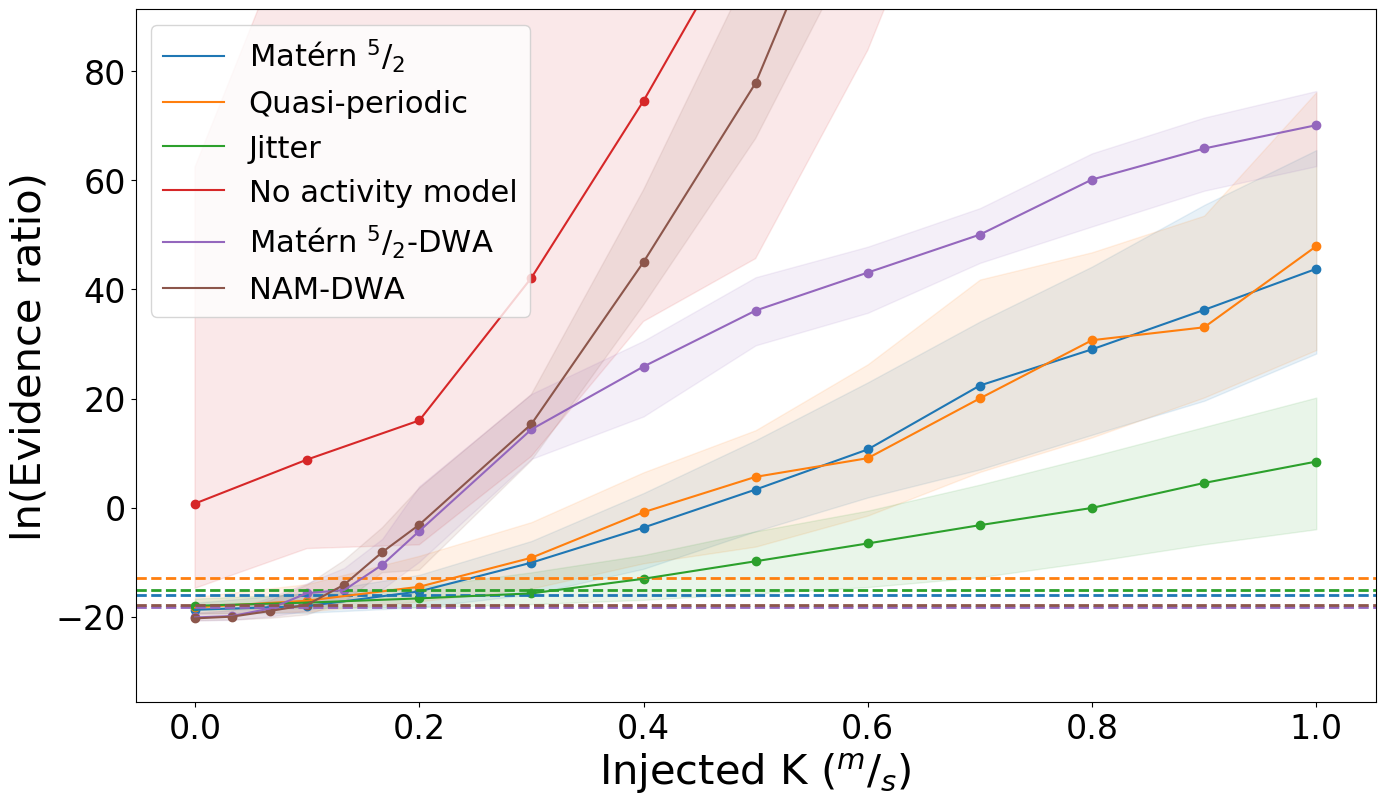}
\caption{The distribution of estimated log evidence ratio comparing models with or without a planet on a Keplerian orbit as a function of injected velocity semi-amplitude, $K$. 
The points represent the median log evidence ratio at each tested $K$ and the fill is between the 16\% and 84\% quantiles. 
The horizontal dashed lines are the 95\% quantiles for the evidence ratios of each of the models at $K=0$ m/s (the vertical dashed lines from Fig. \ref{fig:erhist}).
(The 95\% quantile in the no activity model (red) case is not shown and is ln(evidence ratio)$\approx$200.)
One would expect to detect a planet with a 5\% false discovery rate at least half of the time for planets with $K$ equal or greater than the location where the solid curve crosses the corresponding (shared color) dashed line.}
\label{fig:eroverK}
\end{figure}

Next, we quantify the planet detection sensitivity of each of the above models by analyzing simulated data sets after adding a Keplerian signal to the apparent radial velocities. 
Each injected planetary signal has an orbital period $P=$sqrt(72)$\approx$8.485 days, $0<e<0.2$, 
% TODO: Were e's drawn from uniform or Rayleigh distribution?
and random orientation angles. 
We explore a range of velocity amplitudes from $K=0-1$ m/s, generating 50 simulated data sets for each $K$, each containing 100 observations randomly selected from one year of simulated data from \citetalias{Gilbertson2020}.
The S/N and analysis procedure is the same as for the 0-planet data sets in Fig. \ref{fig:erhist}.
In Fig. \ref{fig:eroverK}, we summarize the resulting distribution of the evidence ratios as a function of $K$ by showing the median log evidence ratio at each tested $K$ (points) and the 16\% and 84\% quantiles (boundaries of filled regions). 
These should be compared to the horizontal dashed lines of the same color located at the 95\% quantiles for the evidence ratios of each of the models at $K=0$ m/s (the vertical dashed lines from Fig. \ref{fig:erhist}). 
One would expect to detect a planet with a 5\% false discovery rate at least half of the time for planets with $K$ equal or greater than the location where the solid curve crosses the corresponding dashed line (same color).

\begin{figure}
\centering
\includegraphics[width=18cm]{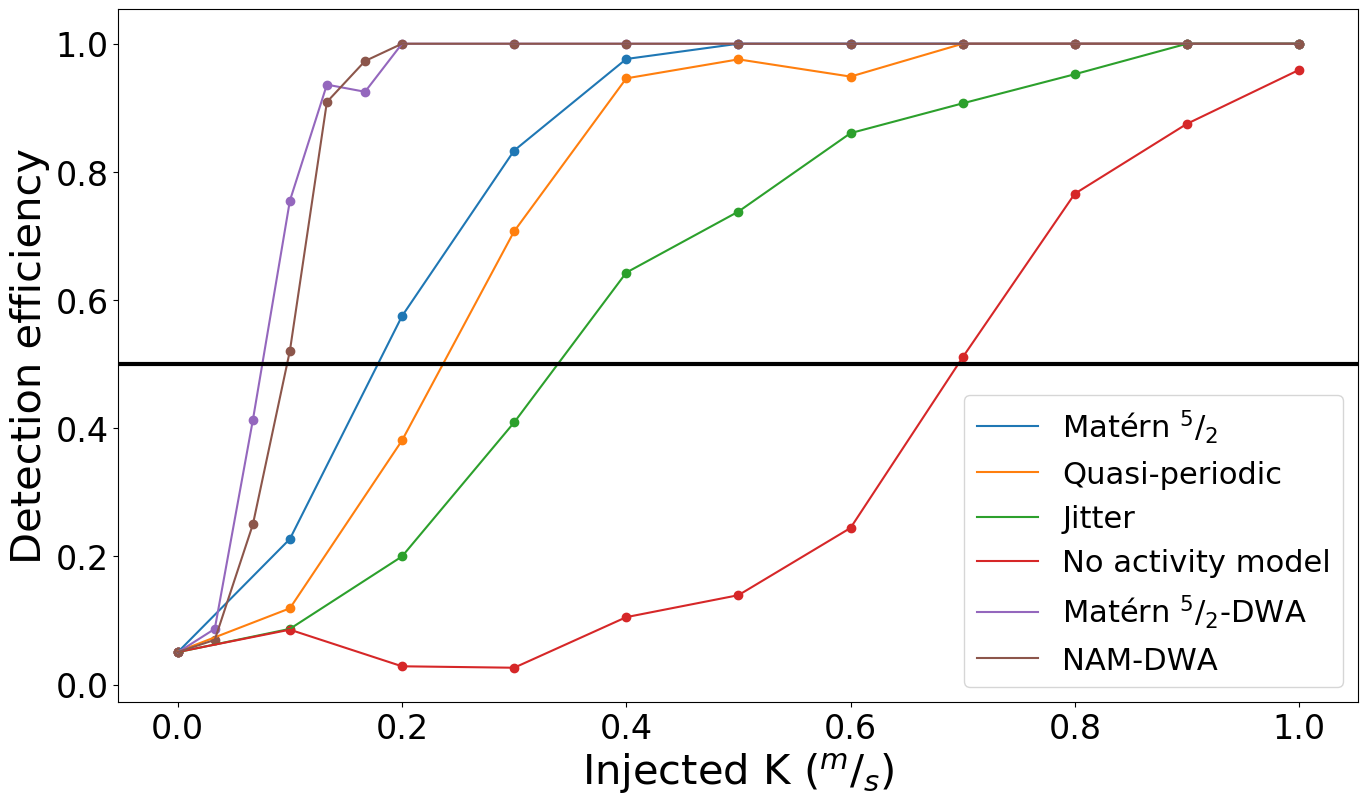}
\caption{The fraction of planets ``detected'' as a function of their velocity amplitude, $K$, for various stellar activity models. 
Each point is based on analyzing 50 data sets that include a Keplerian signal with $P\approx$8.485 days, $0<e<0.2$, and randomized angles. 
The detection criterion is that the evidence ratio (for 1-planet model relative to the 0-planet model) exceed a critical evidence ratio based on the 
95\% quantile for the same evidence ratio based on analyzing similar data sets with no planetary signal (see Fig.\ \ref{fig:erhist}). 
As expected, the detection sensitivity is maximized if the data are generated without any stellar activity (brown and purple curves).
For all other curves, the simulated data includes solar-like stellar activity from \citepalias{Gilbertson2020}. 
Ignoring stellar activity (red) or treating it as uncorrelated white noise (green) results in substantially reduced sensitivity compared to using a \glom model with either a QP (orange) or M$^5/_2$ (blue) kernel. 
For parameters of our simulations, the M$^5/_2$ kernel outperforms the more commonly used QP kernel.}
\label{fig:detections}
\end{figure}

%Taking the 95\% quantile for these distributions allows us to get a heuristic for what evidence ratio would be considered anomalously high for any of our activity models when a planet isn't present.
Finally, we compare the power of different models for detecting low-mass planets in the presence of stellar variability. 
The detection efficiency of a given model and injected $K$ is defined as the proportion of the evidence ratios that exceeded a critical threshold (Fig. \ref{fig:detections}). 
Here the critical threshold for evidence ratio to yield a detection was set based on the 95\% quantile of the evidence ratio distribution computed for 0-planet case (vertical dashed lines in Fig. \ref{fig:erhist}). 
%\ebf{maybe we need subsection so we can refer to them} \cjg{or just referring to the figure when they first appear?}
Unsurprisingly, the sensitivity for detecting low-mass planets would be maximized if there were no stellar variability (brown and purple lines). 
Of course, this is not realistic for most solar-like stars.
At the other extreme, ignoring stellar activity (red) or treating it as jitter (green) result in significantly lower sensitivity than the \glom model using either a M$^5/_2$ (blue) or QP (orange) kernel. 
In this analysis, the \glom models using either QP or M$^5/_2$ kernels allow for the detection of planets roughly half as massive as if stellar activity were treated as white noise or jitter. 
While the evidence ratio distributions for the \glom models using the QP and M$^5/_2$ kernels track very closely, the detection efficiencies for the QP kernel are reduced due to its greater variance in the evidence ratio for the 0-planet case. 
This is likely due to the relatively short lifetimes of the sunspots \citepalias[with a median lifetime of $\approx$ 1 day;][]{Gilbertson2020} which prevents most spots from appearing for multiple stellar rotations.

%Using the evidence ratio threshold allows us to probe the ability of each model to distinguish RVs from stellar activity from true Keplerian signals.

\section{Discussion} \label{sec:discussion}

We have presented a new tool for modelling multivariate time series where each observable is modeled as a linear combination of a latent GP and its derivatives. 
We have demonstrated the \glom statistical model and \glomjl software package to analyzing a time series of RVs and activity indicators calculated from a set of empirically-informed solar spectra.
We show that the \glom model for jointly modeling a planet and stellar variability significantly outperforms either ignoring stellar variability or treating it as white noise. 
For the test cases considered, we show that the \glom model with the Mat\'ern $^5/_2$ GP kernel provides more sensitivity to low-mass planets than the frequently used quasiperiodic kernel. 
We anticipate that \glomjl can accelerate future research to improve the sensitivity of EPRV surveys, such as exploring more GP kernels or the utility of various activity indicators (either traditional or new indicators based on data-driven methods).
Since \glomjl supports a class of models, it can be used in power-based stellar activity model selection, a general strategy posed by \citetalias{Jones2017}, and illustrated here. Statistical techniques for stellar activity model selection are a key focus for improving exoplanet detection.
It can also be incorporated into more detailed models that allow for practical complications such as telluric absorption, observing with multiple instruments, or searching for multiple planets around one star. 

In the future, we look forward to further improvements to the \glomjl package. For example, it could be accelerated for large data sets by speeding up matrix factorizations, perhaps making use of HODLR structure in the covariance matrix \citep[e.g. george;][]{Ambikasaran2014}, or by incorporating special, computationally-efficient, kernel functions \citep[e.g.,]{Foreman-Mackey2017, Loper2020}. 

\acknowledgments
% TODO: will have to edit this as the other authors have been added
The authors wish to thank Tom Loredo, Jessica Cisewski, Hyungsuk Tak, Jinglin Zhao, and Robert Wolpert for useful discussions. 
% The authors wish to thank Ari Silburt for writing the initial versions of the SOAP 2.0 scripts used in this work and Xavier Dumusque for customizing SOAP 2.0 for our purposes. 
C.G. acknowledges the support of the Pennsylvania State University and Pennsylvania Space Grant Consortium. 
This work was funded in part by NSF AST award \#1616086 and NASA Exoplanets Research Program \#80NSSC18K0443.
This research was supported by Heising-Simons Foundation Grant \#2019-1177 (E.B.F.). 
This work was supported by a grant from the Simons Foundation/SFARI (675601, E.B.F.).
E.B.F. acknowledges the support of the Ambrose Monell Foundation and the Institute for Advanced Study.
%
%This research has made use of the NASA Exoplanet Archive, which is operated by the California Institute of Technology, under contract with the National Aeronautics and Space Administration under the Exoplanet Exploration Program.
%
C.G. and E.B.F. acknowledge support from the Penn State Eberly College of Science and Department of Astronomy \& Astrophysics, the Center for Exoplanets and Habitable Worlds and the Center for Astrostatistics. 
The citations in this paper have made use of NASA's Astrophysics Data System Bibliographic Services. 
We acknowledge the Institute for Computational and Data Sciences (\url{http://icds.psu.edu/}) at The Pennsylvania State University, including the CyberLAMP cluster supported by NSF grant MRI-1626251, for providing advanced computing resources and services that have contributed to the research results reported in this paper.
This study benefited from the 2016/2017 Program on Statistical, Mathematical and Computational Methods for Astronomy, and their associated working groups.
This material was based upon work partially supported by the National Science Foundation under Grant DMS-1127914 to the Statistical and Applied Mathematical Sciences Institute (SAMSI). Any opinions, findings, and conclusions or recommendations expressed in this material are those of the author(s) and do not necessarily reflect the views of the National Science Foundation.
%
%\vspace{5mm}
\software{Julia \citep{Bezanson2017}, Optim.jl \citep{Mogensen2018}, SymEngine.jl \citep{Meurer2017}, \added{GPLinearODEMaker.jl \citep{GilbertsonZenodo}}}

%\newpage
\appendix

\section{Hyperparameter priors} \label{sec:priors}

To find models that most closely align with both the data and our prior beliefs, we maximize the log unnormalized posterior, $\ell(\theta|\textbf{x}) + \text{ln}(p(\theta))$. 
For the coefficient hyperparameters, $a_{ij}$, we adopt an unbounded, uniform prior %(\ref{eq:uniform}) 
and therefore $p(\theta) \propto p(\phi)$.
% \begin{equation}
% p_{Uniform}(\lambda) \propto 1
%  \label{eq:uniform}
% \end{equation}
%
%\noindent 
A uniform distribution is the maximum entropy probability distribution for a continuous quantity whose the mean and variance are not known, as is the case for $a_{ij}$. 
In practice, this does not appear to create a problem with improper posteriors, at least for our test applications. 
If one wanted to be certain that posteriors would remain proper, then one could replace the uniform prior on $a_{ij}$'s with a Gaussian distributions with zero mean and large variance. 
For the SE and M$^5/_2$ %, and PP 
kernels we use a gamma distribution as the prior on the single length scale, $\lambda$ (\ref{eq:gamma}).

\begin{equation}
 p_{Gamma}(\lambda, \alpha, \beta) = \dfrac{1}{\Gamma(\alpha) \beta^\alpha} \lambda^{\alpha-1}\text{exp}(-\dfrac{\lambda}{\beta}),
 \label{eq:gamma}
\end{equation}

\noindent where $\alpha$ is the shape parameter and $\beta$ is the scale parameter. 
The gamma prior was chosen for its desirable boundary conditions (0 at both 0 and $\infty$), unimodality, and smoothness.
%For our modeling of the stellar activity described in \S\ref{sec:ex_app}, 
We set $\alpha$ = 5.83 and $\beta$ = 6.21 days, so that the prior mode was 30 days and the standard deviation of the prior was 15 days, motivated by {\em a priori} expectation that correlation length scales are likely to be of the order of the stellar rotation period. 
% When $X(t)$ was constructed using a second SE or M$^5/_2$ based GP, we included another gamma prior on the second $\lambda$ where $\alpha$ and $\theta$ led to a mode and standard deviation 60 and 15 days respectively (i.e. $\alpha$ = 17.94, $\theta$ = 3.54 days).
%
For the QP kernel, a bivariate normal distribution was used as the prior on $\boldsymbol{x} = \{\lambda_{SE}, \tau_P\}$ with mean $\mu_{\tau_P}= 60$ days, standard deviations $\sigma_{\lambda_{SE}} = \sigma_{\tau_P} = 15$ days and the correlation between the two $\rho = 0.9$, resulting in a covariance matrix 
$\Sigma= \begin{pmatrix}
\sigma_{\lambda_{SE}}^2 & \rho \ \sigma_{\lambda_{SE}} \ \sigma_{\tau_P} \\
\rho \ \sigma_{\lambda_{SE}} \ \sigma_{\tau_P} & \sigma_{\tau_P}^2 \\
\end{pmatrix}$.
We have found that, in practice, fitting the quantity $\dfrac{1}{\lambda_P}$ was more stable than fitting for $\lambda_P$.
Therefore, we used a gamma prior on $\dfrac{1}{\lambda_P}$. 
We set $\alpha = 8.13$ days$^{-1}$ and $\beta = 0.14$ days$^{-1}$, so that that the prior mode is 1 day$^{-1}$, and the prior standard deviation is 0.4 day$^{-2}$, yielding a prior that favors similar contributions from the squared exponential and periodic components. 
The \glom model priors are summarized in Tab. \ref{tab:glomprior}.

\begin{table}[ht]
\begin{tabular}{|l|l|l|l|}
\hline
\glom parameter & Kernel & Prior distribution & Prior parameters \\ \hline
$a_{i,j}$ & All & Uniform, unbounded & None \\ \hline
$\lambda_{SE}$ & SE & Gamma & $\alpha$ = 5.83, $\beta$ = 6.21 days \\ \hline
$\lambda_{\text{M}^5/_2}$ & M$^5/_2$ & Gamma & $\alpha$ = 5.83, $\beta$ = 6.21 days \\ \hline
$\lambda_{SE}, \tau_P$ & QP & Bivariate normal & $\mu_{\lambda_{SE}}$= 30 days, $\mu_{\tau_P}$= 60 days, $\sigma_{\lambda_{SE}} = \sigma_{\tau_P} = 15, \ \rho = 0.9$ \\ \hline
$\dfrac{1}{\lambda_P}$ & QP & Gamma & $\alpha$ = 8.13, $\beta$ = 0.14 days$^{-1}$ \\ \hline
\end{tabular}
\caption{A summary of the \glom model priors used in the analysis described in \S\ref{sec:ex_app}.}
\label{tab:glomprior}
\end{table}

\section{Keplerian priors} \label{sec:keppriors}

For models with a planet, we adopt broad priors for the Keplerian model parameters, as summarized in Tab. \ref{tab:kepprior}.
In order to enable Bayesian model comparison, all priors for model parameters for planets must be proper. 
Therefore, we truncate the prior distributions based on a combination of mathematical and physical considerations.
For the planet velocity semi-amplitude, $K$, and period, $P$, we start with a noninformative Jeffreys prior for scale parameters, i.e., uniform in logarithm of the parameter, before truncating to allow for proper normalization,
\begin{equation}
 p_{Logarithmic}(\lambda; \lambda_{\max}, \lambda_{\min}) = \dfrac{1}{\ln\left(\lambda_{\max}/\lambda_{\min}\right)\lambda}, \;\; \lambda_{\min} \le \lambda \le \lambda_{\max}.
 \label{eq:logarithmic}
\end{equation}
\noindent 
%The $K$ and $P$ priors were respectively bound to the intervals [$10^{-4}$ m/s, 2129 m/s] and [1 day, 1000 years].
%
For the planet eccentricity, $e$, prior, we start with a Rayleigh distribution with $\sigma_e=0.2$ and truncate it to have support only over interval [0, 1) (\ref{eq:rayleigh}).
\begin{equation}
 p(e; \sigma) = \dfrac{e}{C\sigma^2} \text{exp}\left[-e^2/(2\sigma^2)\right], \;\; 0 \le e < 1
 \label{eq:rayleigh}
\end{equation}
where $C = 1-e^{-1/\left(2\sigma^{2}\right)}$ is the corresponding cumulative distribution evaluated at the upper limit to ensure proper normalization.
For the planet's argument of periapsis, $\omega$, and initial mean anomaly, $M_0$, we adopt a uniform prior on the interval [0, 2$\pi$], based on geometric considerations. % (\ref{eq:uniform}).
For the radial velocity offset term, $\gamma$ a uniform prior on the interval [-2129 m/s, 2129 m/s] was adopted.

\begin{table}[ht]
\begin{tabular}{|l|l|l|l|}
\hline
Keplerian parameter & Base Prior distribution & Minimum & Maximum \\ \hline
$K$, Velocity semi-amplitude & Logarithmic & $10^{-4}$ m/s & 2129 m/s \\ \hline
$P$, Period & Logarithmic & 1 day & 1000 years \\ \hline
$M_0$, Initial mean anomaly & Uniform & 0 & 2$\pi$ \\ \hline
$e$, Eccentricity & Rayleigh($\sigma_e=0.2$) & 0 & 1 \\ \hline
$\omega$, Argument of periapsis & Uniform & 0 & 2$\pi$ \\ \hline
$\gamma$, Radial velocity offset & Uniform & -2129 m/s & 2129 m/s \\ \hline
\end{tabular}
\caption{A summary of the Keplerian model priors used in the analysis described in \S\ref{sec:ex_app}.}
\label{tab:kepprior}
\end{table}

\section{Sample code} \label{sec:samplecode}

We offer sample code demonstrating how one can create a \glom model using \glomjlns.
In this example, we fit a bivariate time series where the two components are generated from sine and cosine curves, respectively.
Further documentation can be found at the GitHub repository, \href{https://github.com/christiangil/GPLinearODEMaker.jl}{https://github.com/christiangil/GPLinearODEMaker.jl}.\\

%\begin{minted}{julia}
\begin{tt}
\# sample script \\
import GPLinearODEMaker \\
GLOM = GPLinearODEMaker \\
kernel, n\_kern\_hyper = include("../src/kernels/se\_kernel.jl") \# made with kernel\_coder() \\ \\

\# creating some data\\
n = 100 \\
xs = 20 .* sort(rand(n)) \\
noise1 = 0.1 .* ones(n) \\ 
noise2 = 0.2 .* ones(n) \\
y1 = sin.(xs) .+ (noise1 .* randn(n)) \\
y2 = cos.(xs) .+ (noise2 .* randn(n)) \\
ys = collect(Iterators.flatten(zip(y1, y2))) \# putting the outputs in a list, one observation at a time \\
noise = collect(Iterators.flatten(zip(noise1, noise2))) \# putting the errors in a list, one observation at a time \\
n\_dif = 1 + 1 \# base case + 1 differentiation order \\
n\_out = 2 \# 2 dimensional model \\ \\

\# creating a model (or problem definition) with the first output as a GP and the second as a derivative of the base GP \\
prob\_def = GLOM.GLO(kernel, n\_kern\_hyper, n\_dif, n\_out, xs, ys; noise = noise, a0=[[1. 0];[0 1]]) \\ \\

\# making a list with all of the model hyperparameters, including an initial guess for $\lambda_{SE}$ \\
total\_hyperparameters = append!(collect(Iterators.flatten(prob\_def.a0)), [10]) \\ \\

\# initializing a workspace to hold values reused in the optimization \\
workspace = GLOM.nlogL\_matrix\_workspace(prob\_def, total\_hyperparameters) \\

\# creating single-input functions for the likelihood/posterior, gradient, and Hessian

\# optimizing on the posterior requires adding priors to these functions \\
function f(non\_zero\_hyper::Vector\{T\} where T<:Real) = GLOM.nlogL\_GLOM!(workspace, prob\_def, non\_zero\_hyper) \\ \\
function g!(G::Vector{T}, non\_zero\_hyper::Vector{T}) where T<:Real

G[:] = GLOM.$\nabla$nlogL\_GLOM!(workspace, prob\_def, non\_zero\_hyper) \\
end \\
function h!(H::Matrix{T}, non\_zero\_hyper::Vector{T}) where T<:Real

  H[:, :] = GLOM.$\nabla\nabla$nlogL\_GLOM!(workspace, prob\_def, non\_zero\_hyper) \\
end \\ \\

\# the initial guess for the non-zero model parameters \\
initial\_x = GLOM.remove\_zeros(total\_hyperparameters) \\ \\

\# performing optimization using Optim \\
using Optim \\
\# result = optimize(f, initial\_x, NelderMead()) \# slow or wrong \\
\# result = optimize(f, g!, initial\_x, LBFGS()) \# faster and usually right \\
result = optimize(f, g!, h!, initial\_x, NewtonTrustRegion()) \# fastest and usually right
\end{tt}
% \end{minted}

\bibliography{bibliography}

\begin{thebibliography}{}
\expandafter\ifx\csname natexlab\endcsname\relax\def\natexlab#1{#1}\fi
\providecommand{\url}[1]{\href{#1}{#1}}
\providecommand{\dodoi}[1]{doi:~\href{http://doi.org/#1}{\nolinkurl{#1}}}
\providecommand{\doeprint}[1]{\href{http://ascl.net/#1}{\nolinkurl{http://ascl.net/#1}}}
\providecommand{\doarXiv}[1]{\href{https://arxiv.org/abs/#1}{\nolinkurl{https://arxiv.org/abs/#1}}}

\bibitem[{{Aigrain} {et~al.}(2015){Aigrain}, {Hodgkin}, {Irwin}, {Lewis}, \&
  {Roberts}}]{Aigrain2015}
{Aigrain}, S., {Hodgkin}, S.~T., {Irwin}, M.~J., {Lewis}, J.~R., \& {Roberts},
  S.~J. 2015, \mnras, 447, 2880, \dodoi{10.1093/mnras/stu2638}

\bibitem[{{Aigrain} {et~al.}(2012){Aigrain}, {Pont}, \& {Zucker}}]{Aigrain2012}
{Aigrain}, S., {Pont}, F., \& {Zucker}, S. 2012, \mnras, 419, 3147,
  \dodoi{10.1111/j.1365-2966.2011.19960.x}

\bibitem[{{Ambikasaran} {et~al.}(2015){Ambikasaran}, {Foreman-Mackey},
  {Greengard}, {Hogg}, \& {O'Neil}}]{Ambikasaran2014}
{Ambikasaran}, S., {Foreman-Mackey}, D., {Greengard}, L., {Hogg}, D.~W., \&
  {O'Neil}, M. 2015, IEEE Transactions on Pattern Analysis and Machine
  Intelligence, 38, 252, \dodoi{10.1109/TPAMI.2015.2448083}

\bibitem[{{Anglada-Escude} {et~al.}(2014){Anglada-Escude}, {Arriagada},
  {Tuomi}, {Zechmeister}, {Jenkins}, {Ofir}, {Dreizler}, {Gerlach}, {Marvin},
  {Reiners}, {Jeffers}, {Butler}, {Vogt}, {Amado}, {Rodriguez-Lopez},
  {Berdinas}, {Morin}, {Crane}, {Shectman}, {Thompson}, {Diaz}, {Rivera},
  {Sarmiento}, \& {Jones}}]{AngladaEscude2014}
{Anglada-Escude}, G., {Arriagada}, P., {Tuomi}, M., {et~al.} 2014, \mnras, 443,
  L89, \dodoi{10.1093/mnrasl/slu076}

\bibitem[{{Angus} {et~al.}(2018){Angus}, {Morton}, {Aigrain}, {Foreman-Mackey},
  \& {Rajpaul}}]{Angus2018}
{Angus}, R., {Morton}, T., {Aigrain}, S., {Foreman-Mackey}, D., \& {Rajpaul},
  V. 2018, \mnras, 474, 2094, \dodoi{10.1093/mnras/stx2109}

\bibitem[{Bezanson {et~al.}(2017)Bezanson, Edelman, Karpinski, \&
  Shah}]{Bezanson2017}
Bezanson, J., Edelman, A., Karpinski, S., \& Shah, V.~B. 2017, SIAM Review, 59,
  65, \dodoi{10.1137/141000671}

\bibitem[{{Collier Cameron} {et~al.}(2019){Collier Cameron}, {Mortier},
  {Phillips}, {Dumusque}, {Haywood}, {Langellier}, {Watson}, {Cegla}, {Costes},
  {Charbonneau}, {Coffinet}, {Latham}, {Lopez-Morales}, {Malavolta},
  {Maldonado}, {Micela}, {Milbourne}, {Molinari}, {Saar}, {Thompson},
  {Buchschacher}, {Cecconi}, {Cosentino}, {Ghedina}, {Glenday}, {Gonzalez},
  {Li}, {Lodi}, {Lovis}, {Pepe}, {Poretti}, {Rice}, {Sasselov}, {Sozzetti},
  {Szentgyorgyi}, {Udry}, \& {Walsworth}}]{CollierCameron2019}
{Collier Cameron}, A., {Mortier}, A., {Phillips}, D., {et~al.} 2019, \mnras,
  487, 1082, \dodoi{10.1093/mnras/stz1215}

\bibitem[{{Davis} {et~al.}(2017){Davis}, {Cisewski}, {Dumusque}, {Fischer}, \&
  {Ford}}]{Davis2017}
{Davis}, A.~B., {Cisewski}, J., {Dumusque}, X., {Fischer}, D.~A., \& {Ford},
  E.~B. 2017, \apj, 846, 59, \dodoi{10.3847/1538-4357/aa8303}

\bibitem[{{Dumusque} {et~al.}(2014){Dumusque}, {Boisse}, \&
  {Santos}}]{Dumusque2014}
{Dumusque}, X., {Boisse}, I., \& {Santos}, N.~C. 2014, \apj, 796, 132,
  \dodoi{10.1088/0004-637X/796/2/132}

\bibitem[{{Dumusque} {et~al.}(2012){Dumusque}, {Pepe}, {Lovis},
  {S{\'e}gransan}, {Sahlmann}, {Benz}, {Bouchy}, {Mayor}, {Queloz}, {Santos},
  \& {Udry}}]{Dumusque2012}
{Dumusque}, X., {Pepe}, F., {Lovis}, C., {et~al.} 2012, \nat, 491, 207,
  \dodoi{10.1038/nature11572}

\bibitem[{{Dumusque} {et~al.}(2017){Dumusque}, {Borsa}, {Damasso}, {D{\'\i}az},
  {Gregory}, {Hara}, {Hatzes}, {Rajpaul}, {Tuomi}, {Aigrain},
  {Anglada-Escud{\'e}}, {Bonomo}, {Bou{\'e}}, {Dauvergne}, {Frustagli},
  {Giacobbe}, {Haywood}, {Jones}, {Laskar}, {Pinamonti}, {Poretti}, {Rainer},
  {S{\'e}gransan}, {Sozzetti}, \& {Udry}}]{Dumusque2017}
{Dumusque}, X., {Borsa}, F., {Damasso}, M., {et~al.} 2017, \aap, 598, A133,
  \dodoi{10.1051/0004-6361/201628671}

\bibitem[{Duvenaud(2014)}]{Duvenaud2014}
Duvenaud, D. 2014, PhD thesis, University of Cambridge

\bibitem[{{Feroz} \& {Hobson}(2014)}]{FerozHobson2014}
{Feroz}, F., \& {Hobson}, M.~P. 2014, \mnras, 437, 3540,
  \dodoi{10.1093/mnras/stt2148}

\bibitem[{{Fischer} {et~al.}(2016){Fischer}, {Anglada-Escude}, {Arriagada},
  {Baluev}, {Bean}, {Bouchy}, {Buchhave}, {Carroll}, {Chakraborty}, {Crepp},
  {Dawson}, {Diddams}, {Dumusque}, {Eastman}, {Endl}, {Figueira}, {Ford},
  {Foreman-Mackey}, {Fournier}, {F{\H{u}}r{\'e}sz}, {Gaudi}, {Gregory},
  {Grundahl}, {Hatzes}, {H{\'e}brard}, {Herrero}, {Hogg}, {Howard}, {Johnson},
  {Jorden}, {Jurgenson}, {Latham}, {Laughlin}, {Loredo}, {Lovis}, {Mahadevan},
  {McCracken}, {Pepe}, {Perez}, {Phillips}, {Plavchan}, {Prato}, {Quirrenbach},
  {Reiners}, {Robertson}, {Santos}, {Sawyer}, {Segransan}, {Sozzetti},
  {Steinmetz}, {Szentgyorgyi}, {Udry}, {Valenti}, {Wang}, {Wittenmyer}, \&
  {Wright}}]{Fischer2016}
{Fischer}, D.~A., {Anglada-Escude}, G., {Arriagada}, P., {et~al.} 2016,
  Publications of the Astronomical Society of the Pacific, 128, 066001,
  \dodoi{10.1088/1538-3873/128/964/066001}

\bibitem[{{Ford}(2006)}]{Ford2006}
{Ford}, E.~B. 2006, \apj, 642, 505, \dodoi{10.1086/500802}

\bibitem[{{Foreman-Mackey} {et~al.}(2017){Foreman-Mackey}, {Agol},
  {Ambikasaran}, \& {Angus}}]{Foreman-Mackey2017}
{Foreman-Mackey}, D., {Agol}, E., {Ambikasaran}, S., \& {Angus}, R. 2017, \aj,
  154, 220, \dodoi{10.3847/1538-3881/aa9332}

\bibitem[{Gilbertson {et~al.}(2020{\natexlab{a}})Gilbertson, Ford, \&
  Dumusque}]{Gilbertson2020}
Gilbertson, C., Ford, E.~B., \& Dumusque, X. 2020{\natexlab{a}}, Research Notes
  of the {AAS}, 4, 59, \dodoi{10.3847/2515-5172/ab8d44}

\bibitem[{Gilbertson {et~al.}(2020{\natexlab{b}})Gilbertson, Ford, Jones, \&
  Stenning}]{GilbertsonZenodo}
Gilbertson, C., Ford, E.~B., Jones, D.~E., \& Stenning, D.~C.
  2020{\natexlab{b}}, christiangil/GPLinearODEMaker.jl: Zenodo release, v0.1.3,
   Zenodo, \dodoi{10.5281/zenodo.4144107}

\bibitem[{{Grunblatt} {et~al.}(2015){Grunblatt}, {Howard}, \&
  {Haywood}}]{Grunblatt2015}
{Grunblatt}, S.~K., {Howard}, A.~W., \& {Haywood}, R.~D. 2015, \apj, 808, 127,
  \dodoi{10.1088/0004-637X/808/2/127}

\bibitem[{{Haywood} {et~al.}(2014){Haywood}, {Collier Cameron}, {Queloz},
  {Barros}, {Deleuil}, {Fares}, {Gillon}, {Lanza}, {Lovis}, {Moutou}, {Pepe},
  {Pollacco}, {Santerne}, {S{\'e}gransan}, \& {Unruh}}]{Haywood2014}
{Haywood}, R.~D., {Collier Cameron}, A., {Queloz}, D., {et~al.} 2014, \mnras,
  443, 2517, \dodoi{10.1093/mnras/stu1320}

\bibitem[{{Hu} \& {Tak}(2020)}]{HuTak2020}
{Hu}, Z., \& {Tak}, H. 2020, \aj, 160, 265, \dodoi{10.3847/1538-3881/abc1e2}

\bibitem[{{Jones} {et~al.}(2017){Jones}, {Stenning}, {Ford}, {Wolpert},
  {Loredo}, \& {Dumusque}}]{Jones2017}
{Jones}, D.~E., {Stenning}, D.~C., {Ford}, E.~B., {et~al.} 2017, ArXiv
  e-prints.
\newblock \doarXiv{1711.01318}

\bibitem[{{Littlefair} {et~al.}(2017){Littlefair}, {Burningham}, \&
  {Helling}}]{Littlefair2017}
{Littlefair}, S.~P., {Burningham}, B., \& {Helling}, C. 2017, \mnras, 466,
  4250, \dodoi{10.1093/mnras/stw3376}

\bibitem[{{Liu} {et~al.}(2018){Liu}, {Jiang}, {Huang}, {Yu}, {Yang}, {Jia},
  {Awiphan}, {Pan}, {Liu}, {Zhang}, {Wang}, {Li}, {Du}, {Li}, {Lu}, {Zhang},
  {Tian}, {Li}, {Ji}, {Zhang}, {Shi}, {Wang}, {Zhou}, \& {Zhou}}]{Liu2018}
{Liu}, H.-G., {Jiang}, P., {Huang}, X., {et~al.} 2018, \aj, 155, 12,
  \dodoi{10.3847/1538-3881/aa9b86}

\bibitem[{{Loper} {et~al.}(2020){Loper}, {Blei}, {Cunningham}, \&
  {Paninski}}]{Loper2020}
{Loper}, J., {Blei}, D., {Cunningham}, J.~P., \& {Paninski}, L. 2020, arXiv
  e-prints, arXiv:2003.05554.
\newblock \doarXiv{2003.05554}

\bibitem[{{Lovis} \& {Fischer}(2010)}]{Lovis2010}
{Lovis}, C., \& {Fischer}, D. 2010, {Radial Velocity Techniques for
  Exoplanets}, ed. S.~{Seager} (University of Arizona Press Tucson, AZ), 27--53

\bibitem[{Meurer {et~al.}(2017)Meurer, Smith, Paprocki, \v{C}ert\'{i}k,
  Kirpichev, Rocklin, Kumar, Ivanov, Moore, Singh, Rathnayake, Vig, Granger,
  Muller, Bonazzi, Gupta, Vats, Johansson, Pedregosa, Curry, Terrel,
  Rou\v{c}ka, Saboo, Fernando, Kulal, Cimrman, \& Scopatz}]{Meurer2017}
Meurer, A., Smith, C.~P., Paprocki, M., {et~al.} 2017, PeerJ Computer Science,
  3, e103, \dodoi{10.7717/peerj-cs.103}

\bibitem[{Mogensen \& Riseth(2018)}]{Mogensen2018}
Mogensen, P.~K., \& Riseth, A.~N. 2018, The Journal of Open Source Software, 3,
  615, \dodoi{10.21105/joss.00615}

\bibitem[{{Nava} {et~al.}(2020){Nava}, {L{\'o}pez-Morales}, {Haywood}, \&
  {Giles}}]{Nava2020}
{Nava}, C., {L{\'o}pez-Morales}, M., {Haywood}, R.~D., \& {Giles}, H. A.~C.
  2020, \aj, 159, 23, \dodoi{10.3847/1538-3881/ab53ec}

\bibitem[{Nelson {et~al.}(2020)Nelson, Ford, Buchner, Cloutier, D{\'{\i}}az,
  Faria, Hara, Rajpaul, \& Rukdee}]{Nelson2020}
Nelson, B.~E., Ford, E.~B., Buchner, J., {et~al.} 2020, The Astronomical
  Journal, 159, 73, \dodoi{10.3847/1538-3881/ab5190}

\bibitem[{{Pepe} {et~al.}(2002){Pepe}, {Mayor}, {Galland}, {Naef}, {Queloz},
  {Santos}, {Udry}, \& {Burnet}}]{Pepe2002}
{Pepe}, F., {Mayor}, M., {Galland}, F., {et~al.} 2002, \aap, 388, 632,
  \dodoi{10.1051/0004-6361:20020433}

\bibitem[{{Rajpaul} {et~al.}(2015){Rajpaul}, {Aigrain}, {Osborne}, {Reece}, \&
  {Roberts}}]{Rajpaul2015}
{Rajpaul}, V., {Aigrain}, S., {Osborne}, M.~A., {Reece}, S., \& {Roberts}, S.
  2015, \mnras, 452, 2269, \dodoi{10.1093/mnras/stv1428}

\bibitem[{{Rajpaul} {et~al.}(2016){Rajpaul}, {Aigrain}, \&
  {Roberts}}]{Rajpaul2016}
{Rajpaul}, V., {Aigrain}, S., \& {Roberts}, S. 2016, \mnras, 456, L6,
  \dodoi{10.1093/mnrasl/slv164}

\bibitem[{{Rasmussen} \& {Williams}(2006)}]{Rasmussen2006}
{Rasmussen}, C.~E., \& {Williams}, C. K.~I. 2006, {Gaussian Processes for
  Machine Learning} (MIT Press Cambridge, MA)

\bibitem[{{Robertson} \& {Mahadevan}(2014)}]{Robertson2014b}
{Robertson}, P., \& {Mahadevan}, S. 2014, \apjl, 793, L24,
  \dodoi{10.1088/2041-8205/793/2/L24}

\bibitem[{{Robertson} {et~al.}(2014){Robertson}, {Mahadevan}, {Endl}, \&
  {Roy}}]{Robertson2014a}
{Robertson}, P., {Mahadevan}, S., {Endl}, M., \& {Roy}, A. 2014, Science, 345,
  440, \dodoi{10.1126/science.1253253}

\bibitem[{{Robertson} {et~al.}(2015){Robertson}, {Roy}, \&
  {Mahadevan}}]{Robertson2015}
{Robertson}, P., {Roy}, A., \& {Mahadevan}, S. 2015, \apjl, 805, L22,
  \dodoi{10.1088/2041-8205/805/2/L22}

\bibitem[{{Thompson} {et~al.}(2020){Thompson}, {Watson}, {Haywood}, {Costes},
  {de Mooij}, {Collier Cameron}, {Dumusque}, {Phillips}, {Saar}, {Mortier},
  {Milbourne}, {Aigrain}, {Cegla}, {Charbonneau}, {Cosentino}, {Ghedina},
  {Latham}, {L{\'o}pez-Morales}, {Micela}, {Molinari}, {Poretti}, {Sozzetti},
  {Thompson}, \& {Walsworth}}]{Thompson2020}
{Thompson}, A.~P.~G., {Watson}, C.~A., {Haywood}, R.~D., {et~al.} 2020, \mnras,
  494, 4279, \dodoi{10.1093/mnras/staa1010}

\bibitem[{{Vogt} {et~al.}(2010){Vogt}, {Butler}, {Rivera}, {Haghighipour},
  {Henry}, \& {Williamson}}]{Vogt2010}
{Vogt}, S.~S., {Butler}, R.~P., {Rivera}, E.~J., {et~al.} 2010, \apj, 723, 954,
  \dodoi{10.1088/0004-637X/723/1/954}

\bibitem[{{Wright} \& {Howard}(2009)}]{Wright2009}
{Wright}, J.~T., \& {Howard}, A.~W. 2009, \apjs, 182, 205,
  \dodoi{10.1088/0067-0049/182/1/205}

\bibitem[{{Zechmeister} {et~al.}(2018){Zechmeister}, {Reiners}, {Amado},
  {Azzaro}, {Bauer}, {B{\'e}jar}, {Caballero}, {Guenther}, {Hagen}, {Jeffers},
  {Kaminski}, {K{\"u}rster}, {Launhardt}, {Montes}, {Morales}, {Quirrenbach},
  {Reffert}, {Ribas}, {Seifert}, {Tal-Or}, \& {Wolthoff}}]{Zechmeister2018}
{Zechmeister}, M., {Reiners}, A., {Amado}, P.~J., {et~al.} 2018, \aap, 609,
  A12, \dodoi{10.1051/0004-6361/201731483}

\end{thebibliography}
\bibliographystyle{aasjournal}

\end{document}